\documentclass[aip,preprint]{revtex4-1}
\usepackage{amsmath}
\usepackage{graphicx}
\usepackage{multirow}
\usepackage{epstopdf}
\draft 

\begin{document}

\title{Dissipative particle dynamics simulations of a single isolated polymer chain in a dilute solution} 

\author{Praphul Kumar}
\author{Harishyam}
\author{Indranil Saha Dalal}
\email[Author to whom correspondence should be 
addressed; electronic mail: ]{indrasd@iitk.ac.in}
\affiliation{Indian Institute of Technology Kanpur, Kanpur-208016, India}


\date{\today}
\begin{abstract}
In this study, we investigate the suitability of dissipative particle dynamics (DPD) simulations to predict the dynamics of polymer chains in dilute polymer solutions, where the chain is represented by a set of beads connected by almost inextensible springs. In terms of behaviour, these springs closely mimic rods that serve as representations of Kuhn steps. We find that the predictions depend on the value of the repulsive parameter for bead-bead pairwise interactions used in the DPD simulations ($a_{ij}$). For all systems, the chain sizes and the relaxation time spectrum are analyzed. For $a_{ij} = 0$, theta solvent behaviour is obtained for the chain size, whereas the dynamics at equilibrium agrees well with the predictions of the Zimm model. For higher values of $a_{ij}$, the static properties of the chain show good solvent behaviour. However, the scaling laws for the chain dynamics at equilibrium show wide variations, with consistent results obtained only at an intermediate value of $a_{ij} = 25$. At higher values of the repulsive parameter ($a_{ij}  \geq 25$), our simulations are also able to predict the abrupt cut-off in the relaxation spectrum, which has been observed earlier in experiments of dilute solutions. The cut-off reached an extent that, for chain lengths of 10 Kuhn steps, the spectrum consists of a single time scale. This agrees remarkably well with earlier experiments and MD simulations. To verify further, we also studied the chain dynamics in shear flow using DPD simulations. Specifically, we analysed the variation of the chain stretch and end-over-end tumbling with shear rates. Overall, the trends obtained from DPD simulations agree well with those observed in earlier BD simulations.
\end{abstract}

\maketitle 
\section{introduction}

The knowledge of the dynamics of polymer chains in solution are of enormous importance for the prediction of various rheological properties such as diffusivity, viscosity etc. It is well established that the properties are linked with the conformational changes of the polymer chains at microscopic length scales. In this regard, Rouse \cite{rouse1953theory} developed the first micro-mechanical model to capture the dynamics of polymer chain in dilute solutions using the normal mode analysis. In his model, the polymer chain is constructed by a string of beads connected by Hookean springs, and considered the forces on the beads due to springs, drag and the Brownian force due to the thermal motion of the solvent. However, he ignored the effect of hydrodynamic interactions (HI), which arises due to the movement of beads influencing the dynamics of all other beads. Rouse obtained the scaling laws for chain diffusion coefficient as $D \sim N^{-1}$ and the chain relaxation time as $\tau \sim N^2$, where $N$ represents the number of beads. Later, Zimm \cite{zimm1956dynamics} added a correction to the Rouse model by adding the effect of HI in a pre-averaged manner, and predicted the scaling laws as $D \sim N^{- \nu}$ and $\tau \sim N^{3\nu}$, where $\nu$ is the Flory's exponent. The value of $\nu$ for good, bad and theta solvent is 3/5, 1/3 and 1/2, respectively. Experiments confirmed that the predictions of the Zimm model agrees well with the observations in dilute polymer solutions.

Over the years, computer simulations have emerged as a great tool to explain the microscopic chain dynamics in polymer solutions at equilibrium and under an imposed flow field. The results from Brownian dynamics (BD) simulations of bead-rod and bead-spring model for polymer chains, with and without HI, in shear flows \cite{hur2000brownian}, correctly captures the trends observed in DNA single-molecule imaging experiment \cite{smith1999single} in shear flow. In this approach, the solvent is treated as a continuum and hence, reduces the large number of degrees of freedom associated with the solvent molecules. However, a BD simulation incorporating HI effects become computationally expensive beyond a relatively small number of beads. On the other hand, results from molecular dynamics (MD) simulations of polymer chain using an implicit solvent captured the scaling laws predicted by Rouse model but failed to agree well with experiments \cite{kaznessis1998molecular}, which is expected since HI is neglected in these simulations. Therefore, it is imperative to use a simulation method that can correctly incorporate the effects of HI. MD simulations have also been performed with explicit solvent
\cite{dunweg1993molecular,polson2006equilibrium}, where HI is implicitly present due to the  solvent molecules in the system. However, this requires the presence of an enormous number of degrees of freedom. Additionally, the requirement of a very small time-step size (typical in MD simulations, for convergence) makes it computationally prohibitive, even at this age of advanced processors.

In addition to these aforementioned approaches, the dissipative particle dynamics (DPD), a relatively new mesoscopic computational method, has drawn attention of researchers and is steadily gaining popularity for studying complex fluids and soft materials [!!!!add ref of recent Review paper !!!!!]. Hoogerbrugge and Koelman \cite{hoogerbrugge1992simulating} were the first to develop the DPD technique, which was modified to its present form by Warren and Espanol \cite{espanol1995statistical}.
DPD simulations have been used in a wide variety of problems such as spinodal decomposition \cite{groot1997dissipative}, nanocomposites \cite{laradji2004nanospheres}, solvent flow through polymer brush, \cite{huang2006flow,wijmans2002simulating} etc. In many such  problems, the interactions at the microscale are important to predict the final structure and dynamics. However, simulations like MD will be able to capture the properties of only  small system sizes at practical timescales. 

The features of DPD are similar to MD, in which a set of soft spheres move according to Newton's law of motion due to pairwise forces. It treats the solvent particles explicitly and hence, is expected to incorporate the HI implicitly between the beads.  The typical interactions between a pair of DPD beads consist of soft repulsive forces, Brownian forces and dissipative forces. Additionally, spring forces will also be  present due to connectors in a polymer chain. The soft repulsive interactions allow a relatively large time-step size for integrating the equations of motion compared to typical MD simulations. The details about the nature of forces are discussed later in this article. DPD simulations were performed earlier for polymer solutions using bead-spring models \cite{jiang2013accurate}. However, recent BD simulations \cite{dalal2012multiple,dalal2014effects} have shown significant differences between the predictions of bead-spring and bead-rod models for an imposed flow field, even at the steady state. Thus, it becomes imperative to study the corresponding behaviour of bead-rod chains, where the solvent molecules are treated explicitly, as in DPD simulations. This study performs detailed DPD simulations of polymer solutions using bead-rod models and tries to ascertain the suitability of the DPD method to simulate a bead-rod chain in a solvent bath, with and without an imposed shear flow. Note here that, by a ``rod", we mean a stiff, almost inextensible spring, which mimics the behaviour of a single Kuhn step of a polymer chain. Such a check for DPD simulations is extremely important owing to known problems of this method. Firstly, the Schimidt number is low, which is not correct for a liquid phase. Secondly, all EV interactions in conventional DPD is handled via soft potentials. In earlier BD simulations \cite{dalal2014effects}, those were modeled by Lennard-Jones potentials, which diverges sharply at short distances. THus, it becomes imperative to check whether all scaling laws of polymer dynamics are reproduced by conventional DPD simulations.

Besides the issue of the discretization of a polymer chain, there have been surprising experimental evidences of low stretch of chains in shear flow for good solvents\cite{lee1999flow}. Surprisingly, the chains showed extensions for a poor solvent but almost no stretch for one good solvent. A clear explanation of these results are not found in literature, to the best of our knowledge. This indicates some lack of understanding of the role of the dynamics of the surrounding solvent molecules when the chain is exposed to a flow field. Issues like this cannot be addressed by BD simulations, where the solvent is replaced by a continuum. In a DPD, the bath of solvent molecules is treated explicitly. Thus, for further investigations into the effects on the chain dynamics induced by that of the solvent molecules and given the fact that MD simulations are computationally prohibitive, a technique like DPD is likely to be highly suitable. 

In this article, we will primarily focus on the dynamics of polymer chains in a solution predicted by DPD simulations. In this study, a single polymer chain, modelled by a series of beads connected by rods, is immersed in a large simulation box filled with free DPD beads that represent the solvent bath, to mimic a dilute solution. As mentioned earlier, even though the DPD method has been used by researchers in a variety of areas, it has never been investigated if the same is able to satisfactorily capture all the scaling laws obtained from the Zimm model. In this study, we will check the validity of DPD simulations to capture the known features of the dynamics of a polymer chain. Note that, we will explore this dynamics with and without an imposed shear flow. Additionally, we will also search the parameter space for pairwise interactions that can appropriately describe the behaviour of dilute polymer solutions. 

This article is organized in various sections. Section II provides the details of the simulation setup and methods employed in DPD simulations. All the results obtained from this method are presented in Section III. Finally, the key findings are summarized in Section IV.

\section{Methodology}
As mentioned earlier, DPD simulations allow us to use intermediate length scales - smaller than the macroscopic  and larger than the atomistic length scales. In this method, a group of atoms or molecules are ``coarse-grained'' into a single unit, called a  ``DPD particle'' or ``bead'', that reduces the large number of degrees of freedom associated with the solvent molecules, resulting in highly increased computational  efficiency. Thus, it neglects the internal motion of the individual solvent  molecules that occur at shorter time scales. These DPD particles influence the motion of  other neighboring DPD particles through pairwise interactions, which vanish after a cut-off distance $r_c$. Unlike the hard sphere potential model where 
the force between the particles become infinity at overlap, DPD 
considers soft potentials and prohibits the force from diverging at overlap. This is logical since the DPD particles are packets of fluid molecules and their centres can overlap as they move through each other.

For such a system, there are three standard forces acting on 
an individual DPD particle. They are the soft repulsive conservative force, the dissipative force and the random force. The soft repulsive conservative force  ensures that the particles remain distributed in  space in accordance with the equilibrium distribution. Due to the ``soft''  nature of this  force, it enables the accessibility of larger time and length scales. The dissipative force is due to drag and is related to the macroscopic viscosity. The random force causes the Brownian motion of the particles. These random forces are uncorrelated and independent of all other particles. The dissipative and random forces balance themselves to form a thermostat that keeps the mean temperature of the system at a constant value.

\subsection{Mathematical formulation}

Consider a system consisting of $N$ DPD particles, each having a 
mass $m$ for simplicity , with  position vectors $\vec{r}_i$ and velocity $\vec{v}_i$. The governing equation of motion of each individual particle can be written by using the Newton's second law of motion as follows:
\begin{equation}\label{eq:1}  
m\frac{d\vec{v}_i}{dt} = \vec{F}_{ij}
\end{equation}
where $\vec{v}_i=d\vec{r}_i/{dt}$ and $\vec{F}_{ij}$ is the total 
inter-particle 
force acting on the $i^{th}$ particle by all other particles. The total force 
$\vec{F}_{ij}$ is given by 
\begin{equation}\label{eq:2}
\vec{F}_{ij}=\vec{F}_{ij}^C +\vec{F}_{ij}^D +\vec{F}_{ij}^R
\end{equation}
where $\vec{F}_{ij}^C$, $\vec{F}_{ij}^D$ and $\vec{F}_{ij}^R$ are the 
soft conservative, dissipative and random forces, respectively. These forces 
are pairwise additive and are given by
\begin{equation}\label{eq:3}
\vec{F}_{ij}^C = w^C (r_{ij}) \hat{r}_{ij}
\end{equation}
\begin{equation}\label{eq:4}
\vec{F}_{ij}^D = -\gamma w^D (r_{ij})(\hat{r}_{ij}\cdot 
\vec{v}_{ij})\hat{r}_{ij}
\end{equation}
\begin{equation}\label{eq:5}
\vec{F}_{ij}^R = \sigma w^R (r_{ij}) \theta_{ij} \hat{r}_{ij}
\end{equation}
where $ \vec{r}_{ij} = \vec{r}_i-\vec{r}_j$,
$\hat{r}_{ij}=\vec{r}_{ij}/|\vec{r}_{ij}|$,
and $ \vec{v}_{ij} = \vec{v}_i-\vec{v}_j$ are the relative position,  
corresponding unit vector and the velocity vector of bead $i$  with respect to bead 
$j$, respectively.  The variables $w^C$, $w^D$ and $w^R$ are the weight 
functions of the conservative, dissipative and random forces, respectively. 
The parameters $\gamma$ and $\sigma$  determine the strengths of the 
dissipative  and  random forces, respectively. The term $\theta_{ij}$ are the 
Gaussian random 
variables with the symmetry property $ \theta_{ij}= \theta_{ji}$, which 
ensures the total conservation of momentum and have the following properties

\begin{equation}\label{eq:6}
\left\langle  \theta_{ij}\right\rangle =0
\end{equation}
\begin{equation}\label{eq:7}
\left\langle  \theta_{ij}(t) \theta_{kl}(t')\right\rangle 
=(\delta_{ik}\delta_{jl}+\delta_{il}\delta_{jk})\delta(t-t')
\end{equation}

All the forces act within a sphere of cut-off radius $r_c$, which is the 
length scale for the interactions. The conservative force is derived from a soft 
potential, and its weight function can be defined as a function of distance as
\begin{equation}\label{eq:8}
w^C(r_{ij})  =
\begin{cases}
a_{ij}(1-r_{ij}/r_c)  &\text{if $ r_{ij}\leq r_c$}	\\
0 &\text{if $r_{ij}\geq r_c$}
\end{cases}
\end{equation}
where $a_{ij}$ is the repulsion parameter between beads $i$ and $j$. 
This repulsion parameter is one of the most important aspects of DPD 
simulations, as will be observed in this study as well. To be consistent with the fluctuation-dissipation theorem, two 
conditions are set on the weight functions and amplitudes of the 
dissipative and random 	forces 
\cite{espanol1995statistical,groot1997dissipative}
\begin{equation}\label{eq:9}
w^D(r_{ij})  = [w^R(r_{ij})]^2
\end{equation} 
\begin{equation}\label{eq:10}
\sigma^2 = 2\gamma k_BT
\end{equation}
where $k_B$ is the Boltzmann constant and T is the system temperature. 
In the standard DPD method, the weight function takes the following form \cite{groot1997dissipative}
\begin{equation}\label{eq:11}
w^R(r_{ij})  =
\begin{cases}
(1-r_{ij}/r_c)  &\text{if $ r_{ij}\leq r_c$}	\\
0 &\text{if $r_{ij}\geq r_c$}
\end{cases}
\end{equation}

The time evolution of the DPD bead, which is described by Eqs. \ref{eq:1} 
and \ref{eq:2}, can be written as: 

\begin{equation}\label{eq:12}
d\vec{v}_i =\frac{1}{m}\left( 
\vec{F}_{ij}^Cdt+\vec{F}_{ij}^Ddt+\vec{F}_{ij}^R\sqrt{dt}\right)
\end{equation}
The $\sqrt{dt}$ term multiplying random force in Eq. \ref{eq:12} 
ensures that the diffusion coefficient of the particles is independent of the 
time step size used in simulations\cite{groot1997dissipative}. Thus, the exact representation of the  
random 	force given in Eq. (\ref{eq:5})	takes the following form 
\begin{equation}\label{eq:13}
\vec{F}_{ij}^R= 
\sigma w^R\left(r_{ij}\right)\frac{\xi_{ij}}{\sqrt{dt}}\hat{r}_{ij}
\end{equation}
where $\xi_{ij}$ is a Gaussian random variable with a  zero mean and unit 
variance.

\subsection{Integration algorithm}
In computer simulations, the trajectories of DPD particles, which is governed by Eq. (\ref{eq:1}), are calculated using numerical integration. Among many available integration schemes like explicit Euler, the Position Verlet algorithm and the Velocity Verlet algorithm, LAMMPS \cite{plimpton1995fast} uses the velocity-Verlet integrator to update the positions and velocities of the DPD particles. We have used LAMMPS for all the DPD simulations performed for this study. Note that, to increase the accuracy, the velocity-Verlet scheme requires a relatively smaller time-step $\Delta t$. The velocity-Verlet algorithm is given as:
\begin{equation}\label{eq:14}
\vec{r}_i(t+\Delta t)=\vec{r}_i(t)+\Delta 
t\vec{v}_i(t)+\frac{(\Delta t)^2}{2m}\vec{F}_i(t)
\end{equation}
\begin{equation}\label{eq:15}
\vec{F}_i(t+\Delta t)=\vec{F}_i\left(\vec{r}_i\left(t+\Delta t\right)\right)
\end{equation} 
\begin{equation}\label{eq:16}
\vec{v}_i(t+\Delta t)= \vec{v}_i(t)+ 
\frac{\Delta t}{2m}\left[\vec{F}_i(t)+\vec{F}_i(t+\Delta t)\right]
\end{equation}
The performance of the integration scheme in DPD can be evaluated by 
monitoring the temporal evolution of the system temperature, radial 
distribution function and other properties.  In our simulations, we 
choose a small time-step that gives a reasonably accurate performance. 
This aspect of the selection of the time-step size is discussed later.

\subsection{Parameters selection}
In this work, we use LJ units to non-dimensionalize all physical quantities of 
interest. For LJ units, the Lennard-Jones potential 
parameters sigma ($\sigma$) and epsilon ($\epsilon$) are taken as units of 
length and energy. LAMMPS \cite{plimpton1995fast}(Large-scale 
Atomic/Molecular Massively 
Parallel 
Simulator) sets these fundamental quantities mass, sigma, 
epsilon, and Boltzmann constant ($k_B$) as unity. All other physical 
quantities are expressed in terms of these fundamental units. The distance, 
time, energy, temperature and pressure are non-dimensionalized by 
$\sigma$, $\left(\dfrac{\epsilon}{m \sigma^2}\right)^{-1/2}$, $\epsilon$, 
$\epsilon/k_B$ and $\epsilon/\sigma^3$, respectively 
\cite{allen2017computer}.

All the simulations are performed in a cubic periodic box. In all 
simulations, we have taken one polymer chain immersed in a bath of 
solvent particles. The box size is taken large enough so that the size of the 
simulation box does not influence the equilibrium radius of gyration of the 
polymer chain. The particle mass $(m)$ , cut-off distance ($r_c$), and 
$k_B$T are taken as unity. 
Following the convention for DPD simulations, the friction coefficient 
$\gamma$ is set to $4.5$\cite{groot1997dissipative}. We have considered 
three different values of the repulsion coefficient $a_{ij}= 0, 
10,	$ and $25$. For some runs, we also take
a 	higher value of $a_{ij}=50$. Each simulation is performed with all 
the three $a_{ij}$ values to check the dependencies of the results on the 
repulsion parameter. The repulsive interactions between DPD particles are 
set equal for all pairs of beads, namely, $a_{ss}=a_{pp}=a_{sp}$, where 
the subscripts $p$ and $s$	denote the polymer and solvent beads, 
respectively, 	and they interact pairwise. The number density $n=3$ is 
fixed for all the DPD simulations.   

We adopt the bead-rod model to 	represent a polymer chain in the DPD simulations. Each polymer bead is represented by a DPD 
particle, 	and consecutive polymer beads are 	connected by a harmonic 
bond described by a potential $E$  given by:
\begin{equation}\label{eq:17}
E =  K({r}-{r_0})^2
\end{equation} 
where $r_0$ is the equilibrium bond distance and $K$ is the spring 
constant including the usual factor of 1/2. We have chosen $r_0$=0.85 for 
the harmonic bond \cite{schlijper1995computer}, and a value of  
$K=5000$, such that it maintains the property of a stiff, nearly inflexible 
rod. An optimum time-step size of $\Delta t=0.001$ is used in the 
simulations, which gives a 	reasonable accuracy. Details of the selection 
of $K$ and $\Delta t$ values are discussed in the following subsection.

\subsection{Selection of the  parameters $K$ and $\Delta t$}

As mentioned earlier, LAMMPS uses the velocity-Verlet integrator to 
update the position and velocity for the next time-step. The velocity-Verlet algorithm has limited accuracy in DPD simulations. This can be overcome by adopting a sufficiently small time-step size, as confirmed by Hafskjold \textit{et al.} 	\cite{hafskjold2004can} and Chaudhri and Lukes 	\cite{chaudhri2010velocity}. However, it increases the computational cost. Therefore, we decide to choose a value of $\Delta t$ such that 
it is reasonably accurate but not computationally prohibitive. For the 
polymer bead-rod model, an appropriate value of $K$ is required to keep 
the bond length fluctuations from the equilibrium length as small as 
possible. 	To select the optimum values of $\Delta t$ and $K$, we perform a  set of simulations with different combinations of $\Delta t$ and $K$ values. In these, we use all the parameters from the study of Schlijper \textit{et al.} \cite{schlijper1995computer} and set $k_BT=1$ for a $10$ bead polymer chain. After running the simulations for the same total time for each combination of $K $ and $\Delta t$, we calculate the bond lengths after every $0.05$ time units.  Probability distributions of bond lengths are calculated for all the combinations. For the value of $K=5000$, the fluctuation in the bond length is very small, about $3\%$ deviation from the mean. We use the results shown in Fig. 1 for the selection of time-step size. We note that, as we decrease the time-step size, then probability distributions of the bond-length shows larger fluctuations away from the equilibrium bond-length ($r_0$). However, to avoid very small 	$\Delta t$ (this incurs a high computational cost), we select the optimum value of $\Delta t=0.001$ and $K=5000$ for our simulations. Using the parameters mentioned above, simulations are run for at least $50$ relaxation times of the polymer chain to obtain good statistics.


\subsection{Chain size and Auto-correlation function(ACF)}

One of the measures of the chain dimension is the root-mean-square 
of the radius of gyration, denoted as $R_g$. 	For beads of equal masses connected by massless 
bonds, the center of mass $\vec{r}_{cm}$ of the chain is given by
\begin{equation}\label{eq:18}
\vec{r}_{cm}=\frac{1}{N}\sum_{i=1}^{N}\vec{r}_i
\end{equation}
where $N$ is the number of beads and $\vec{r}_i$ is position vector of  
the $i^{th}$ bead. $R_g$ is defined as\cite{doi1988theory}
\begin{equation}\label{eq:19}
R_g = \sqrt{\frac{1}{N}\sum_{i=1}^{N} 
\left\langle\left| \vec{r}_i-\vec{r}_{cm}\right| ^2\right\rangle}
\end{equation}
where $\langle .... \rangle$ denotes an ensemble 
average. The $x$ 
component of $R_g$ can be written as
\begin{equation}\label{eq:20}
R_{g,x} = \sqrt{\frac{1}{N}\sum_{i=1}^{N} 
\left\langle\left({x}_i-{x}_{cm}\right)^2\right\rangle}
\end{equation}
where $x_i$ and  $x_{cm}$ denote the $x$-component of the position of the $i^{th}$ bead 
and the center of mass of the chain along the $x$-direction, respectively. 
Similar expressions can be written for $y$ and $z$ components. Using 
similar formulas, $R_{g,y}$ and $R_{g,z}$ can be calculated. In our 
convention for this study, $y$ is the flow direction. The $z$ and $x$ 
directions denote the shear-gradient and vorticity directions, 
respectively.

In our simulations, the radius of gyration is obtained by averaging 
$R_g$ of the polymer chain over a  sufficiently long time after the 
steady state has been reached. 

The auto-correlation function of end-to-end vector of the polymer chain provides an 
estimate of the relaxation time. The end-to-end auto correlation function is defined as\cite{doi1988theory} 

\begin{equation}\label{eq:21}
C(t)=\langle\vec{R}(t)\cdot \vec{R}(0)\rangle
\end{equation}
where $\vec{R}=\vec{r}_N-\vec{r}_1$ is the end-to-end 
vector of the chain. From the end-to-end vector auto-correlation function, 
we can estimate the relaxation time of the chain. Relaxation time is calculated by fitting the auto-correlation function to an exponential decay, as 
given by \cite{doi1988theory} :
\begin{equation}\label{eq:22}
\langle\vec{R}(t)\cdot\vec{R}(0)\rangle \cong \langle\vec{R}^2\rangle 
\exp\left(-\frac{t}{\tau}\right)
\end{equation}
where $\tau$ is the relaxation time.

The autocorrelation of the end-to-end vector does not give a clear picture of the local dynamics of the chain \cite{jain2008effects}. Since most of the end-to-end ACF is expected to fit well with single exponential, it does not indicate the total active modes needed to describe the dynamics. The ACF of the bond vectors on the chain will be a much better indicator of the local modes in dynamics. The bond vector ACF is similar to that of the end-to-end vector ACF with contribution from all the modes, given as\cite{jain2008effects}:
\begin{equation}\label{eq:22a}
\langle \vec{u}.\vec{u}\rangle =\frac{1}{N_s}\sum_{i=1,3,5...}^{N_s}exp(-t/\tau_i)
\end{equation}

where $\tau_i$ is the relaxation time of the $i^{th}$ mode and $N_s$ is the total number of bonds in the chain.

\subsection{Brownian dynamics (BD) simulations}
 In this study, a few BD simulations are also performed for bead-rod 
 and bead-spring polymer models to complement the results of the DPD simulations to check our methods and analysis. We use 
 the same parameter values for $\vec{r}_c, K, \vec{r}_0$ and $k_BT$, as 
 those in the DPD simulations.
 
 In this method, the total force on a particle consists  of a drag force, 
 $\vec{F}_i^d$, on the particle moving through the viscous solvent, a 
 Brownian force $\vec{F}_i^B$ that arises due to random collisions of the bead with the 
 solvent molecules, and other non-hydrodynamic forces 
 $\vec{F}_i^{nh}$. The total force can be written as:
 \begin{equation}\label{eq:22}
 \vec{F}_i^{tot}=\vec{F}_i^d+\vec{F}_i^B+\vec{F}_i^{nh}
 \end{equation}
 This non-hydrodynamic force $\vec{F}_i^{nh}$ includes any external body 
 forces, excluded volume interactions and spring forces. The stochastic 
 differential equation governing the motion of the particle is given by
 \begin{equation}\label{eq:23}
 \frac{d\vec{r}_i}{dt}=\vec{u}_{\infty}(\vec{r}_i)+
 \frac{1}{\zeta}\left[\vec{F}_i^{nh}\left({\vec{r}_i}\right)+
 \vec{F}_i^B(t)\right]
 \end{equation}
 where $\zeta $ is the drag coefficient of an individual bead, $\vec{u}_\infty(\vec{r}_i)$ is the 
 unperturbed velocity of solvent and $\vec{r}_i$ is the positon vector of the $i^{th}$ bead on the polymer chain. The simulations are performed by time integration of these stochastic equations.

\section{Results and discussion}

	We have performed detailed DPD simulations to understand the static and 
	dynamic properties of dilute polymer solutions. 
	Some BD simulations are also performed  to complement our 
	results obtained from the DPD simulations. As stated earlier, we have chosen 
	a cubic periodic box of sufficient length to avoid the effects of the box size.  All 
	the simulations are run for a sufficiently long time, and the properties like 
	radius of gyration, correlation function, relaxation time etc. are calculated 
	after an initial run of $10$ relaxation times (of the chain) so that the system 
	attains equilibrium.  Table \ref{tab1} shows some properties 
	calculated at equilibrium for different values of the repulsive parameter, 
	$a_{ij}$, box-length and number of DPD beads on the polymer chain. We have used four different 
	chain lengths of $ 10, 20, 30$, and $60$ DPD beads, and four different values of the 
	repulsive parameter, $a_{ij}=0, 10, 25$ and $50$.
	
	\begin{table}
		\caption{\label{tab1}Various parameters used and properties calculated at 
			equilibrium with different values of  $a_{ij}$. All quantities are in 
			dimensionless units. Note that, $\tau$ is the relaxation time of the chain defined precisely in the earlier study\cite{dalal2012multiple}.}
		\begin{ruledtabular}
			\begin{tabular}{llllll}
				
				\ttfamily \shortstack{ Repulsive\\parameter} & 
				\ttfamily 
				\shortstack{Box \\ Length} & \ttfamily \shortstack{Number 
					of\\beads(N)} & \ttfamily \shortstack{Total \\ 
					timesteps\\$(\times 10^{-5})$}  & \ttfamily $R_g$ & \ttfamily $\tau$ 
				\\                \hline
				\multirow{4}{*}{$a_{ij}=0$} & \multicolumn{1}{l}{6.41} & 
				\multicolumn{1}{l}{10} & \multicolumn{1}{l}{100} & 
				\multicolumn{1}{l}{1.09} & 
				\multicolumn{1}{l}{3.23}\\\cline{2-6}
				& \multicolumn{1}{l}{9.25} & \multicolumn{1}{l}{20} & 
				\multicolumn{1}{l}{200} & \multicolumn{1}{l}{1.55} & 
				\multicolumn{1}{l}{12.35} \\\cline{2-6}
				& \multicolumn{1}{l}{12.5} & \multicolumn{1}{l}{30} & 
				\multicolumn{1}{l}{300} & \multicolumn{1}{l}{1.89} & 
				\multicolumn{1}{l}{26.50}\\\cline{2-6}
				& \multicolumn{1}{l}{25.0} & \multicolumn{1}{l}{60} & 
				\multicolumn{1}{l}{500} & \multicolumn{1}{l}{2.71} & 
				\multicolumn{1}{l}{80.19}\\\hline
				
				\multirow{4}{*}{$a_{ij}=10$} & \multicolumn{1}{l}{6.41} & 
				\multicolumn{1}{l}{10} & \multicolumn{1}{l}{100} & 
				\multicolumn{1}{l}{0.95} & 
				\multicolumn{1}{l}{3.38}\\\cline{2-6}
				& \multicolumn{1}{l}{9.25} & \multicolumn{1}{l}{20} & 
				\multicolumn{1}{l}{200} & \multicolumn{1}{l}{1.43} & 
				\multicolumn{1}{l}{13.52}\\\cline{2-6}
				& \multicolumn{1}{l}{12.5} & \multicolumn{1}{l}{30} & 
				\multicolumn{1}{l}{300} & \multicolumn{1}{l}{1.83} & 
				\multicolumn{1}{l}{32.19}\\\cline{2-6}
				& \multicolumn{1}{l}{25.0} & \multicolumn{1}{l}{60} & 
				\multicolumn{1}{l}{500} & \multicolumn{1}{l}{2.80} & 
				\multicolumn{1}{l}{126.37}\\\hline
				
				\multirow{4}{*}{$a_{ij}=25$} & \multicolumn{1}{l}{6.41} & 
				\multicolumn{1}{l}{10} & \multicolumn{1}{l}{100} & 
				\multicolumn{1}{l}{0.78}& 
				\multicolumn{1}{l}{3.82}\\\cline{2-6}
				& \multicolumn{1}{l}{9.25} & \multicolumn{1}{l}{20} & 
				\multicolumn{1}{l}{200} & \multicolumn{1}{l}{1.19} & 
				\multicolumn{1}{l}{14.43}\\\cline{2-6}
				& \multicolumn{1}{l}{12.5} & \multicolumn{1}{l}{30} & 
				\multicolumn{1}{l}{300} & \multicolumn{1}{l}{1.53} & 
				\multicolumn{1}{l}{27.59}\\\cline{2-6}
				& \multicolumn{1}{l}{25.0} & \multicolumn{1}{l}{60} & 
				\multicolumn{1}{l}{500} & \multicolumn{1}{l}{2.36} & 
				\multicolumn{1}{l}{106.44}\\\hline
				
				\multirow{4}{*}{$a_{ij}=50$} & \multicolumn{1}{l}{6.41} & 
				\multicolumn{1}{l}{10} & \multicolumn{1}{l}{100} & 
				\multicolumn{1}{l}{0.71}& 
				\multicolumn{1}{l}{6.91}\\\cline{2-6}
				& \multicolumn{1}{l}{9.25} & \multicolumn{1}{l}{20} & 
				\multicolumn{1}{l}{200} & \multicolumn{1}{l}{1.09} & 
				\multicolumn{1}{l}{21.26}\\\cline{2-6}
				& \multicolumn{1}{l}{12.5} & \multicolumn{1}{l}{30} & 
				\multicolumn{1}{l}{300} & \multicolumn{1}{l}{1.39} & 
				\multicolumn{1}{l}{41.48}\\\cline{2-6}
				& \multicolumn{1}{l}{25.0} & \multicolumn{1}{l}{60} & 
				\multicolumn{1}{l}{500} & \multicolumn{1}{l}{2.14} & 
				\multicolumn{1}{l}{118.40}\\ 
			\end{tabular}
		\end{ruledtabular}
	\end{table}
	
	\subsection{Static properties}
	A natural way to characterize the polymer chains is to observe the scaling 
	of the radius of gyration $R_g$ with the number of links. $R_g$ is 
	calculated using Eq. \ref{eq:19}.  Fig. 2 shows the 
	scaling of $R_g$ with the number of rods.
	The scaling exponents for the power law fit of  the $R_g$ is 
	shown in the legend. The exponent obtained for the value of the repulsive 
	parameter $a_{ij}=0$ confirms that the solvent bath behaves as a theta 
	solvent (exponent of about $0.5$), as predicted by Flory 
	\cite{huggins1954principles}. This is expected since there are no 
	excluded 
	volume interactions between the beads on the chain. For all other values of 
	$a_{ij}=10, 25, 50$, the scaling exponent is close to 0.6, which 	
	implies good solvent behavior \cite{huggins1954principles}. This is in 		
	agreement with our expectations, since there is excluded volume 		
	interactions between the beads on the chain, which now resembles a 		
	self-avoiding random walk. However, this needed to be confirmed since DPD simulations use ``soft'' potential between beads, as discussed earlier. Hence, we can conclude that the static properties 
	at equilibrium is in good agreement with theoretical expectations.

\subsection{Dynamic properties at equilibrium}

\subsubsection{Auto-correlation function (ACF)}

The end-to-end vector auto-correlation function shows 
the relaxation dynamics of the chain at equilibrium. Using Eq. \ref{eq:21} and 22, 
we have calculated the relaxation time of the chain for different values of 
$a_{ij}$.   Fig. 3 shows the scaling of relaxation time $\tau$ 
with the number of beads. The scaling exponents for the power law fit of  $\tau$ are 
given  in the legends.

In the dilute regime, since the chain is expected to obey the Zimm model, we expect $\tau\sim N^{3\nu}$. On the 
other hand, if the chain would have obeyed the Rouse model, then 
$\tau\sim N^{1+2\nu}$ would be obtained. Clearly, the scaling exponent computed by our DPD simulations (Fig. 3), for the theta 
solvent case ($a_{ij}=0$) is $1.797 \pm 0.047$, which shows that the dynamics lies between Zimm and Rouse predictions. Since, all our chains are relatively short, the power law exponent may not reach the value of 1.5 as predicted by the Zimm theory.
Then, with the introduction of bead-bead interactions with  
$a_{ij}=10$, the scaling exponent gets closer to the Rouse model. At a higher value of $a_{ij}$, for $a_{ij}=25$, the exponent agrees well with the Zimm model for good solvent ($\nu=0.59$). At an even higher value, for  $a_{ij}=50$, it  approaches $1.5$, which is the 
prediction from Zimm model for theta solvent. These results show that the experimental observations for the scaling of the relaxation time is best recovered from DPD only when $a_{ij}=25$.

Another way to ascertain whether the results are in agreement with the Zimm or the Rouse model is the variation of $\tau$ and $R_g$, as shown in Fig. 4. Zimm model predicts $\tau \sim R_g^3$ 	regardless of the value of $\nu$. However, the Rouse model predicts $\tau 	\sim R_g^{1+1/\nu}$. From the results in Fig. 4, it can be observed that the scaling exponent, for $a_{ij}=25$, is close to $3$, which agrees remarkably well with the Zimm model. For lower values of $a_{ij}$ ($0$ to $10$), it lies in between the predictions of the Zimm and the Rouse model. For a higher value of $a_{ij}$ ($a_{ij}=50$), the exponent is even lower than the predictions of the Zimm model. Thus, from the scaling laws obtained from the relaxation time of the end-to-end vector, we can conclude that $a_{ij}=25$ yields results that are the closest to the predictions of the Zimm model.


As discussed earlier, the end-to-end ACF does not provide the local dynamics of the chain. For this we need to calculate the bond auto-correlation function using Eq. \ref{eq:22a}. The results in Fig. 5 shows that the relaxation of the backbone bonds typically consist of multiple time scales, except for a short chain for higher $a_{ij}$ values of $25$ and $50$. Note that, a single exponential would appear as a straight line in Fig. 5. Any curvature in the ACF, thus, is an indicator of a significant contribution from the other modes. From the plot of $\langle\vec{u}\cdot \vec{u}\rangle$ vs time, we notice that, for a short chain of $N=10$, as the value of $a_{ij}$ is increased, the ACF approaches a single exponential decay, suggesting a single relaxation mode (Fig. 5a). Thus, a relaxation spectrum of a short chain consists of a single time scale.  More precisely, a short chain of 10 beads (9 Kuhn steps), roughly show a single exponential decay, for $a_{ij}\geq25$. This suppression of higher  modes for short polymer chains is surprising, but is consistent with previous experiments \cite{peterson2001apparent} and simulations \cite{jain2008effects,saha2013explaining}.

\subsubsection{Analysis of normal modes}

Normal modes,  introduced first by Rouse \cite{rouse1953theory},   
decouples the otherwise coupled equations of motion (the motion of the $i^{th}$ 
bead depends on the adjacent bead due to the links attached to it) for the 
polymer chain.  This is defined as 
\begin{equation}\label{eq:24}
\vec{q}_i(t) =\frac{1}{N}\sum_{n=1}^{N}\cos(\frac{in\pi}{N}) 
\vec{r}_n(t)
\end{equation}
where $\vec{q}_i(t)$ is the normal coordinate for the $i^{th}$ mode at 
any  time  $t$. The normal modes provide further local details of the chain 
dynamics. Figs. $6a,c,e$ show the variation of 
the relaxation time of the $i^{th}$ mode with $i$ and Figs. $6b,d,f$, with $N/i$ for different values of $a_{ij}$. The relaxation 
time of a normal mode is  obtained by a single exponential fit of the ACF of that mode. The fits are performed for the lower modes, whereas the 
relaxation times for the higher modes saturate. This is expected for chains that 
are finitely discretized.
The relaxation time, $\tau_i$, of the $i^{th}$ mode of a polymer chain of 
length $N$ represents the relaxation of a sub-chain of length $N/i$. The 
higher modes relax faster than the lower modes and the first, or the slowest 
mode, shows the highest relaxation time, as expected.  If 
hydrodynamic interactions are absent, then the relaxation time, $\tau_i$, 
would scale with $i$ according to the Rouse model, i.e. 
$\tau_i\sim(N/i)^{1+2\nu}$ (here $\nu=0.5$ for theta solvent and 
$\nu=0.6$ for good solvent). However, if HI is present, then $\tau_i$ 
would scale with $i$ according to the Zimm model,  i.e., 
$\tau_i\sim(N/i)^{3\nu}$. In our 
DPD simulations, we obtain a scaling exponent of  
approximately $1.724 \pm 0.072$  for $a_{ij}=0$ (theta solvent),  which is 
in between Rouse and Zimm model for theta solvent. As discussed earlier, this is perhaps due to the fact that our chains are relatively short. For $a_{ij}=10$, neither the Rouse nor the 
Zimm model is obeyed perfectly. However, for $a_{ij}=25$, a value of 
$1.861 \pm 0.112$ is obtained, which is consistent with the predictions of 
the Zimm model for a good solvent. Similar scaling 
laws are obtained for the fits of $\tau_i$ vs $N/i$ as well (Figs. 6$b,d,f$). Thus, similar to the conclusions in the previous section, a value of $a_{ij}=25$ yields results that agree well with the predictions of the Zimm model.

	\subsection{BD simulations}

	As mentioned earlier, we have also performed some BD simulations to 	complement the DPD 	results and to 	further check the validity of our 	calculations. Firstly, we perform BD simulations on a chain of beads and 
	nearly inflexible ``rods'', as in our DPD simulation. From the BD simulations 	of chain lengths $N=10$, 30 and 60 beads, we calculate the mode 	relaxation spectrum. Fig. 7$a$ shows the variation of the relaxation time ($\tau_i$) with the mode number ($i$). It is noted that for 	all chain lengths, the  mode relaxation time scales with the same scaling exponent, and $\tau_i$ saturates to a nearly constant value for higher 
	modes. Fig. 7$b$ shows the scaling of $\tau_i$ with 
	$N/i$. The scaling exponent for all chain lengths is approximately same at $1.875 \pm 0.074$. This is close to the Rouse prediction of 2 for a chain in theta solvent without HI. However, we did not get the scaling exponent of $2$, which is expected for this system. This is perhaps due to the fact that we are using harmonic spring with certain equilibrium length, instead of Hookean spring as used  in the Rouse model.
	
	To test this next, we perform  BD simulation with $100$ fene springs, where each spring mimics 400 Kuhn steps. This arrangement takes the system closer to the original  Rouse model, which considers Hookean springs and not rods. This helps us further validate our methods for the calculation of the relaxation time of the normal modes. Since HI is not present, the scaling laws predicted by the Rouse theory for theta solvent is expected. Fig.7$c$ shows the variation of $\tau_i$ with the mode number and Fig. 7$d$ shows the scaling of $\tau_i$ with $N/i$. We obtain a scaling factor of $1.966 \pm 0.06$ for this, which agrees remarkably well with the Rouse model. The overall trends of the variation of $\tau_i$ with $i$ is similar to the behavior obtained from the DPD simulations for the bead-rod model.

\subsection{Polymer dynamics in shear flow}

Here, we use DPD to study polymer chain dynamics under an imposed 
shear flow. The presence of flow results in a complex rheological behavior of the solution. As any flow field is locally linear and any flow near a boundary is approximately a shear flow, it is extremely  important to understand the behavior in this flow. We 
have performed the simulations of bead-rod chains with 	varying number of beads and the DPD repulsion parameter, $a_{ij}$, over a 	wide range of Weissenberg numbers  ($Wi$). Here, the Weissenberg number is defined as:
\begin{equation}
Wi =\dot{\gamma}\tau
\end{equation}
where $\dot{\gamma}$ is the shear rate and $\tau$ is the longest relaxation time of the chain. The calculation of $\tau$ is already discussed earlier in details.  The details of the set-up and the parameters are already discussed in the earlier sections. In this section, we have computed the components of radius of gyration using Eq. \ref{eq:20}.

Figs. 8$a-c$ show the variation of $R_{g,y}$ normalized by its contour length, $L$, as a function of the Weissenberg number, for $a_{ij}= 0, 10, $and $25$, respectively. The normalization is selected in accordance with the earlier study by Saha Dalal \textit{et al.} \cite{dalal2012multiple}. Here, the $y$-component of the radius of gyration gives a measure of the chain size in the flow direction. The solid lines in Fig. 8$a$ approximately  show the different regimes of deformation as discussed by Saha dalal \textit{et al.}\cite{dalal2012multiple}. 
From Fig. 8$a-c$, it can be noticed that the multiple 
deformation regimes of a polymer chain in shear flow, as noted in the earlier studies \cite{link1993light,dalal2012multiple},  also appear in our results from DPD simulations. For small values of $Wi$, $R_{g,y}$ increases and then reaches a plateau and shows a tendency to decrease at very high $Wi$. Similar behavior in shear flow have been observed by  Saha Dalal \textit{et al.} \cite{dalal2012multiple} using BD simulations. The value of $R_{g,y}/L\simeq 0.2$ at the plateau is also consistent with that obtained from earlier BD simulations. Polymer chain of different lengths have different values of $Wi$ for transition from one regime to the other. For higher values of 
$a_{ij}$ i.e. $a_{ij}=10$ or $25$, the chain deformation occur in a similar way as for $a_{ij}=0$. However, the transition from one deformation regime to other takes place at higher values of $Wi$ for $a_{ij}>0$. This can be clearly observed from Figs. $8a-c$. In Figs. $8d-f$, we can show the same results with respect to the shear rate. Here, the universality of the results at high shear rates is observed, which agrees with the trends shown in the study by Saha dalal \textit{et al.}\cite{dalal2012multiple}

For $a_{ij}=0$, all the three deformation regimes can be observed in Fig. 8$a$. It is interesting to note that the chain compression at high Weissenberg numbers i.e. Regime III in the article by Saha Dalal \textit{et al.} \cite{dalal2012multiple}, is also visible in our results. Since, there is no EV for $a_{ij}=0$ but HI is present implicitly, the chain compression at high shear rates is expected in accordance with the observations in the earlier study by Saha dalal \textit{et al.} \cite{dalal2014effects} 
For $a_{ij}=0$, we computed the scaling in regime III (chain compression) for a chain of 60 beads as  
$R_{g,y}$ $\sim$ $Wi^{-0.9336}$. A similar scaling law has been 
observed by Saha Dalal \textit{et al.} \cite{dalal2014effects} using BD simulations, in the presence of HI. For $a_{ij}=10$ and $25$, it is clearly observed  that Regime III  is suppressed. This is expected since, for any positive value of $a_{ij}$, EV would be present within the polymer chain and it will suppress chain compression\cite{dalal2014effects}. Overall, the trends of the chain stretch in shear flow is cosistent with those reported in earlier BD simulations study. \cite{dalal2012multiple,dalal2014effects}.

\subsection{Chain tumbling}

 We also calculated  tumbling times of the end-to-end vector from the DPD simulations for different values of the repulsive parameter $a_{ij}$. In shear flow, the chain experiences equal amounts of extension and rotation. For low $Wi$, the chain remains close to the equilibrium state and behaves approximately as a random coil. As the strength of the flow increases, the chain gets stretched in the flow direction and tumbles over due to the rotational component of the shear flow. The tumbling dynamics and the algorithm to estimate the tumbling time is discussed in great details by Saha Dalal \textit{et.al.} \cite{saha2012tumbling}. We follow the same procedure for the analysis of the tumbling motion.

Figs. $9a-c$ show the variation of the end-to-end tumbling time (normalized by the number of Kuhn steps, $N_k$) with  the Weissenberg number for different chain lengths and for different values of the repulsive parameter ($a_{ij}$). From our results, we observe that, for low values of $Wi$, the tumbling time is almost constant. As $Wi$ increases, the tumbling time shows a power-law decay with respect to $Wi$. For $a_{ij}=0$ (Fig. $9a$),  that represents the theta solvent, the tumbling time approximately scales as $\sim Wi^{-3/4}$, which was also observed by Saha dalal \textit{et al.}\cite{saha2012tumbling} for bead-rod chains with theta solvent. For $a_{ij}=10$, the scaling law for the tumbling time remains the same. However, for $a_{ij}=25$, we observe that  the tumbling time scales as $\sim Wi^{-2/3}$. It is noted in the earlier study of tumbling times \cite{saha2012tumbling} that the exponent of the power-law can vary from $-3/4$ and $-2/3$. For $a_{ij}=0$, even though HI is present, we observe a scaling law of $-3/4$, which is consistent with the observations from BD simulations\cite{saha2012tumbling} for relatively short chains. In the same study, it is noted that the scaling law exponent should become $-2/3$ for dominant HI. Here, we clearly note a scaling law of $-2/3$ for $a_{ij}=25$. Thus, the trends of the tumbling times obtained from DPD simulations agree well with the earlier study. 

\section{Summary}
To summarize, we investigated the dynamics of individual polymer chains in dilute solutions in details through DPD simulations, to understand the suitability of such mesoscale techniques for this problem. We have built chains discretized to the level of a Kuhn step, by using beads connected by nearly inextensible springs that mimic rods. A bead-rod representation is used, instead of beads and springs, owing to the predictions from recent BD simulations \cite{dalal2012multiple} that clearly highlight differences between the predictions obtained from such differences in representation, even for steady shear. Here, we performed an extensive analysis for the chain sizes and dynamics (for all normal modes) at equilibrium, as well as with an imposed shear flow.

We observe that, the results obtained from the DPD simulations show subtle variations with the value of the repulsive parameter ($a_{ij}$) for bead-bead interactions. The variation of the chain size (measured by $R_g$) with the number of rods in the chain shows that the bath behaves as a theta solvent for $a_{ij}=0$ and as a good solvent for any higher value of $a_{ij}$. The scaling of the relaxation time based on the end-to-end vector varies with the value of $a_{ij}$. The predictions of the Zimm model at theta solvent is obtained for $a_{ij}=0$. For $a_{ij}=10$, the scaling law ($\tau\sim N^{1.995}$) is close to the predictions of the Rouse model, rather than that of Zimm model for good solvent. However, this is recovered for intermediate values of $a_{ij}$ ($a_{ij}=25$ in our simulations) and the results agree well with the Zimm model for good solvent as $\tau\sim N^{1.896}$. For even higher values of $a_{ij}$ ($a_{ij}=50$), the scaling law exponent reduces further and gets closer to that of the Zimm model for theta solvent ($\tau\sim N^{1.589}$). Quite similarly, the normal mode analysis shows that for $a_{ij}=10$, a Rouse-like scaling ($\tau_i\sim (N/i)^{1.987}$) is obtained but for $a_{ij}=25$, the exponent agrees with the Zimm model for good solvent ($\tau\sim N^{1.861}$). Further, the DPD simulations for relatively higher values of $a_{ij}$ ($a_{ij}\geq25$) clearly predict an abrupt cut-off in the relaxation spectrum of the chain, which is also observed in earlier experimental studies\cite{peterson2001apparent}. For a short chain of 10 Kuhn steps, the relaxation spectrum is approximately reduced to a single time scale, which is remarkably consistent with experiments and a recent MD simulation.\cite{saha2013explaining}

  To further investigate the appropriateness of DPD simulations, we have performed simulations with an imposed shear flow. The variation of the chain stretch and end-over-end tumbling times are analysed in details for various values of the bead-bead repulsive parameter. Overall, the results are consistent with earlier BD simulations. For $a_{ij}=0$, the trends agree well with those obtained for BD simulations with HI, but without EV. We clearly obtain three regimes of deformation, with chain compression being visible at high shear rates, as observed in earlier BD simulations \cite{sendner2009single,dalal2014effects}. For any higher value of $a_{ij}$, we observe an immediate reduction in the chain compression at high shear rates, while the other two regimes still remain visible. This is also consistent with earlier results\cite{dalal2014effects}, where it was observed that EV suppresses the chain compression, even in the presence of HI. Our simulations without shear flow clearly show that good solvent scaling laws are obtained for higher values of the bead-bead repulsive parameter, which implies a presence of EV.

Thus, the overall behavior obtained from DPD simulations for this problem appear to be in good agreement with those observed in earlier BD simulations, \cite{dalal2012multiple,dalal2014effects} confirming this as a suitable tool for use in such investigations. However, care needs to be taken to with respect to the interaction parameters, which has a profound influence on the final results.

\begin{acknowledgments}
	We wish to acknowledge the generous support from the IIT Kanpur initiation grant and DST Early Career Research Award for arranging the logistics and equipments required to carry out this study. We also acknowledge the HPC center at IIT Kanpur for providing nodes to carry out majorly of these simulations.
	
\end{acknowledgments}

\bibliography{DPD}

\newpage

 \begin{figure}  
 	\includegraphics[width=5.5in,height=5in]{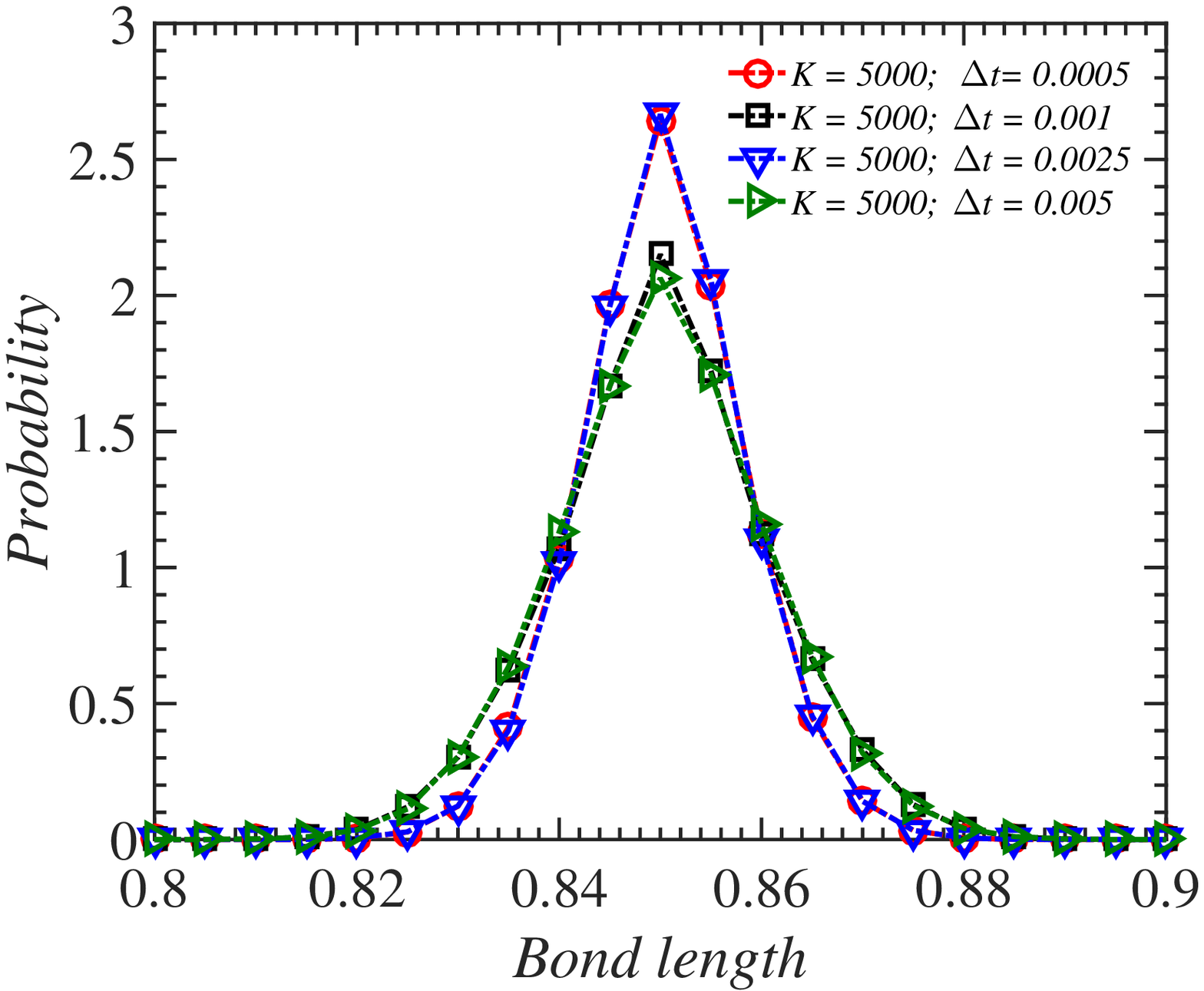} 
 	  \caption {Probaility distribution of the bond length for different time-step sizes.}
 \end{figure}

 \begin{figure}
 	\includegraphics[width=5.5in,height=5in]{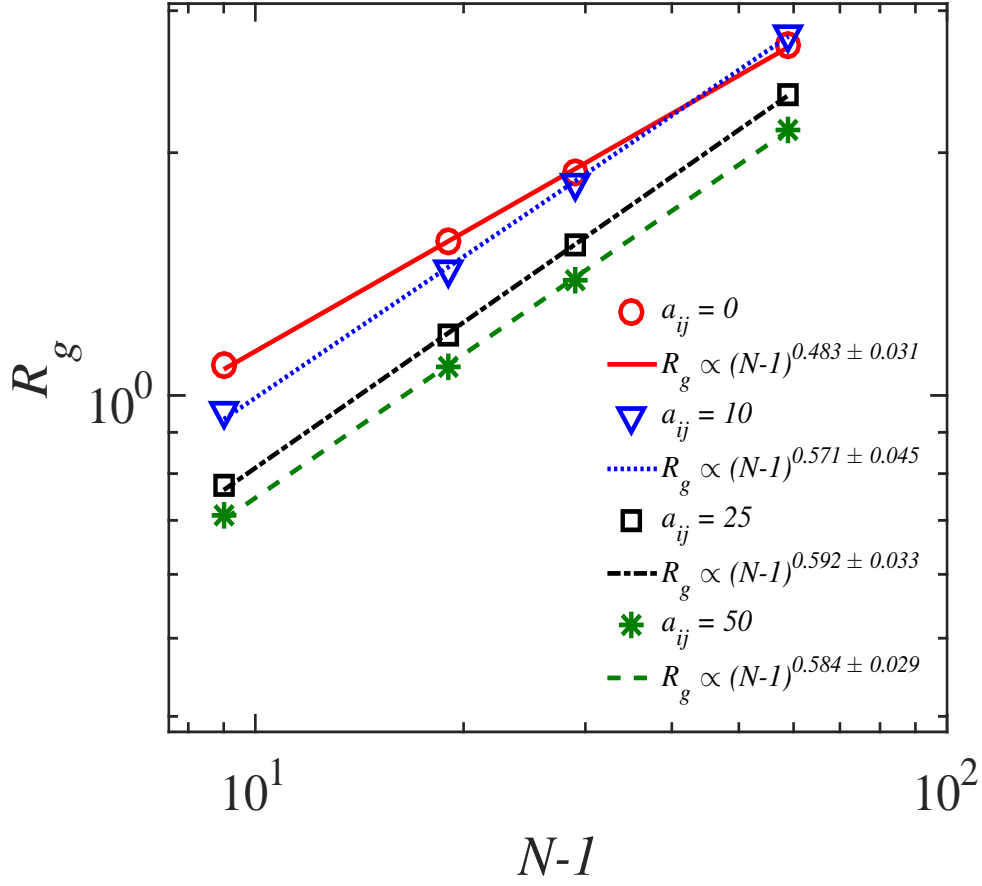}
 	\caption { Variation of $R_g$ with number of rods $(N-1)$, for  different values of repulsive parameter $a_{ij}$. The points represent the  values computed from the DPD simulations, and the corresponding lines are  the power-law fits of the results.}
 \end{figure}
 
 \begin{figure}
 	\includegraphics[width=5.5in,height=5in]{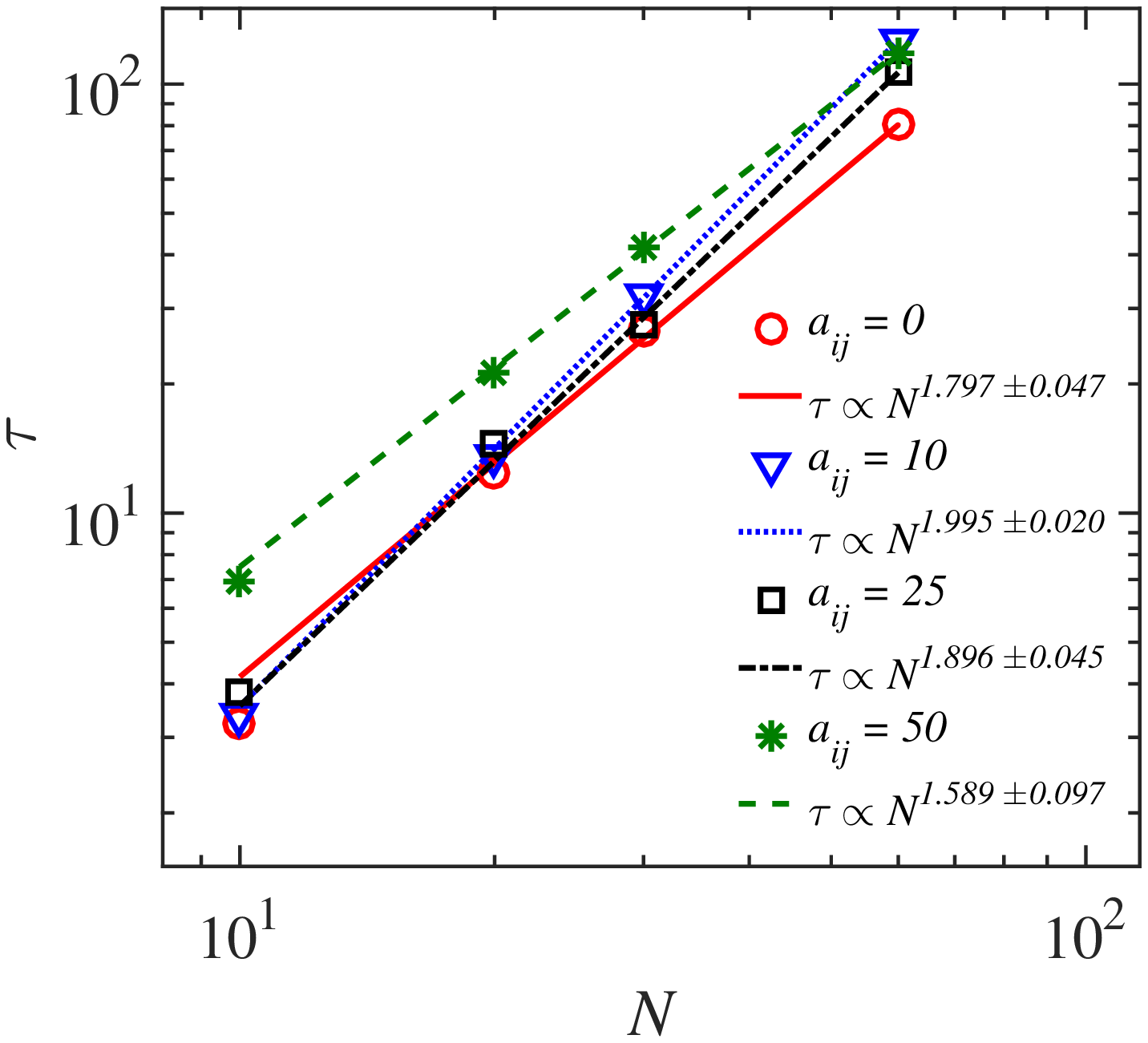}
 	\caption {Variation of the relaxation time $\tau$ with number of beads $N$, for  different values of repulsive parameter $a_{ij}$. The points represent the values computed from the DPD simulations, and the corresponding  lines are the power-law fits of the results.}
 \end{figure}
 
  \begin{figure}
     \includegraphics[width=5.5in,height=5in]{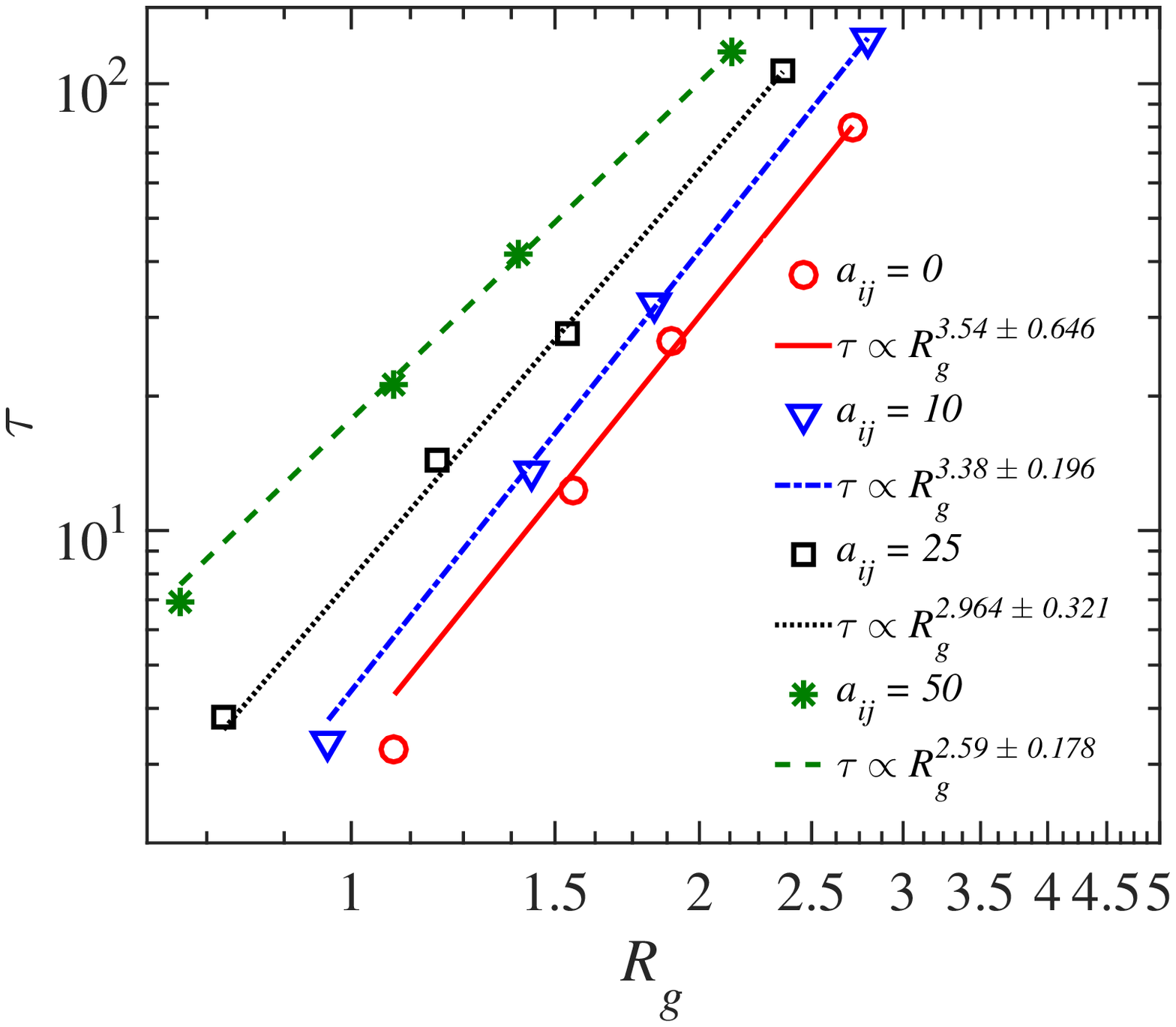}
  	\caption {Variation of the relaxation time $\tau$ with $R_g$, for  different values of repulsive parameter $a_{ij}$. The points represent the values computed from the DPD simulations, and the corresponding lines are the power-law fits of the results.}
  \end{figure}
 
  \begin{figure*}
  	\includegraphics[width=3in,height=2.7in]{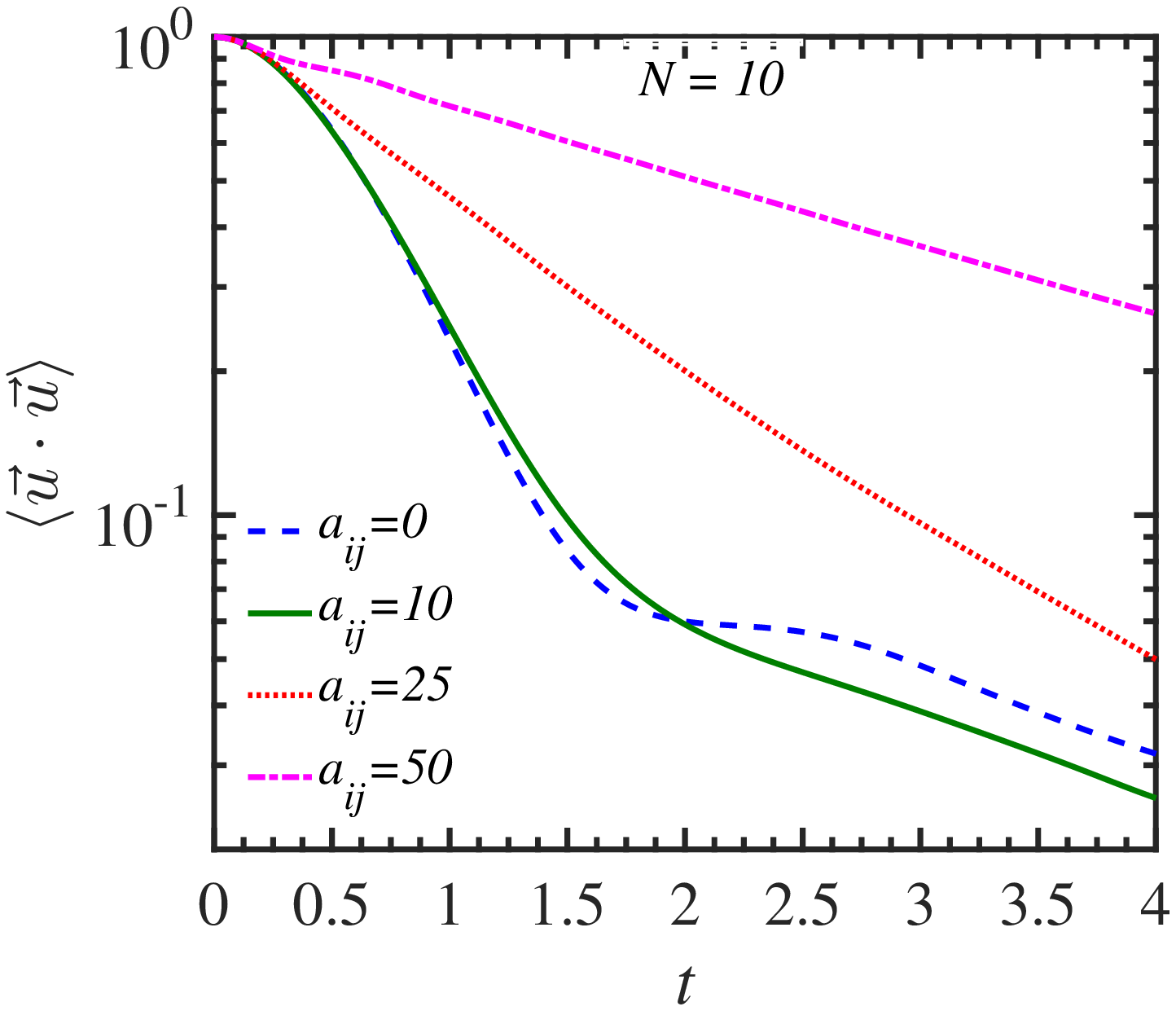}
  	\put (-45,160){($a$)}
  	\includegraphics[width=3in,height=2.7in]{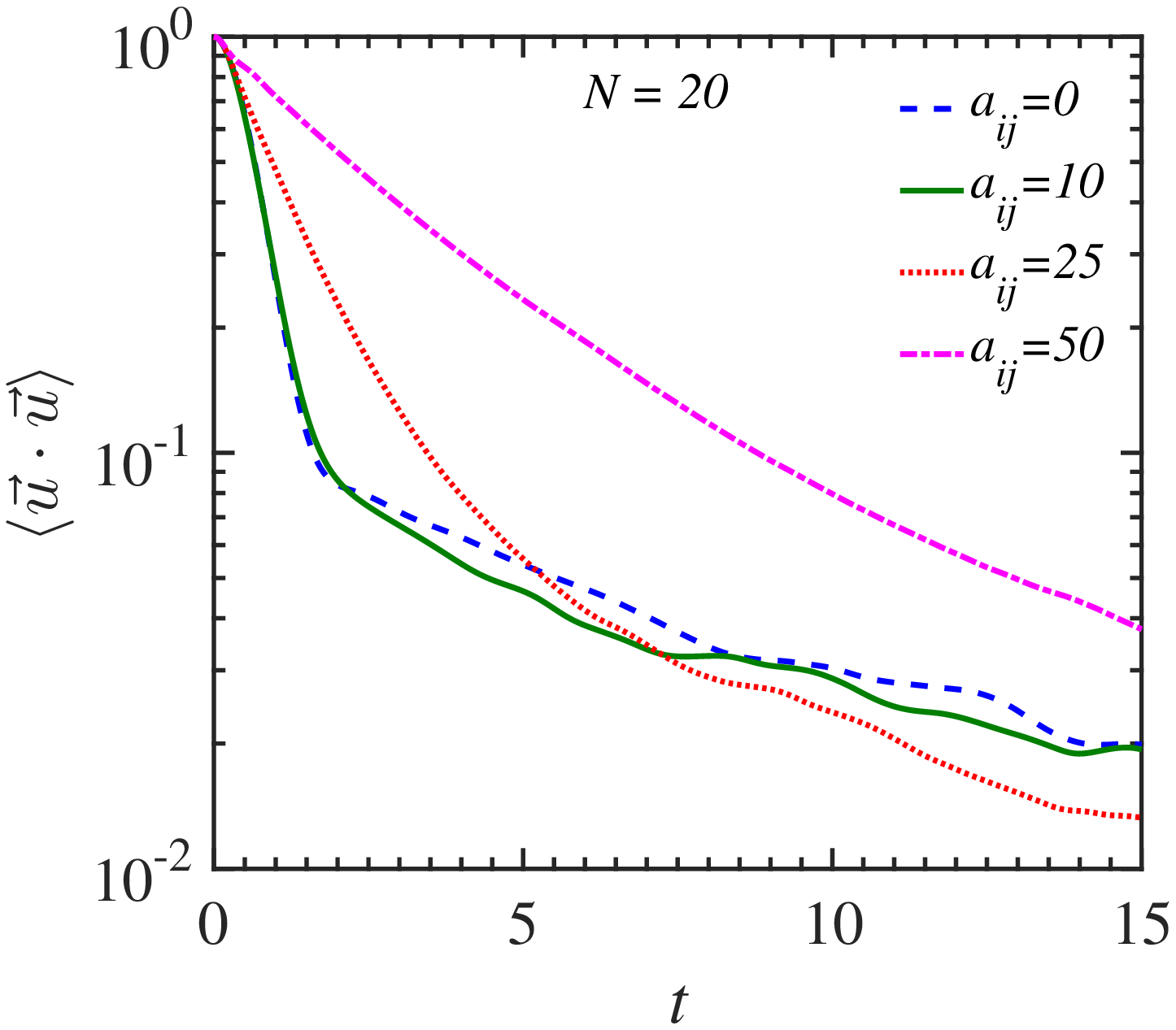}
  	\put (-150,160){($b$)}
  	
  	\includegraphics[width=3in,height=2.7in]{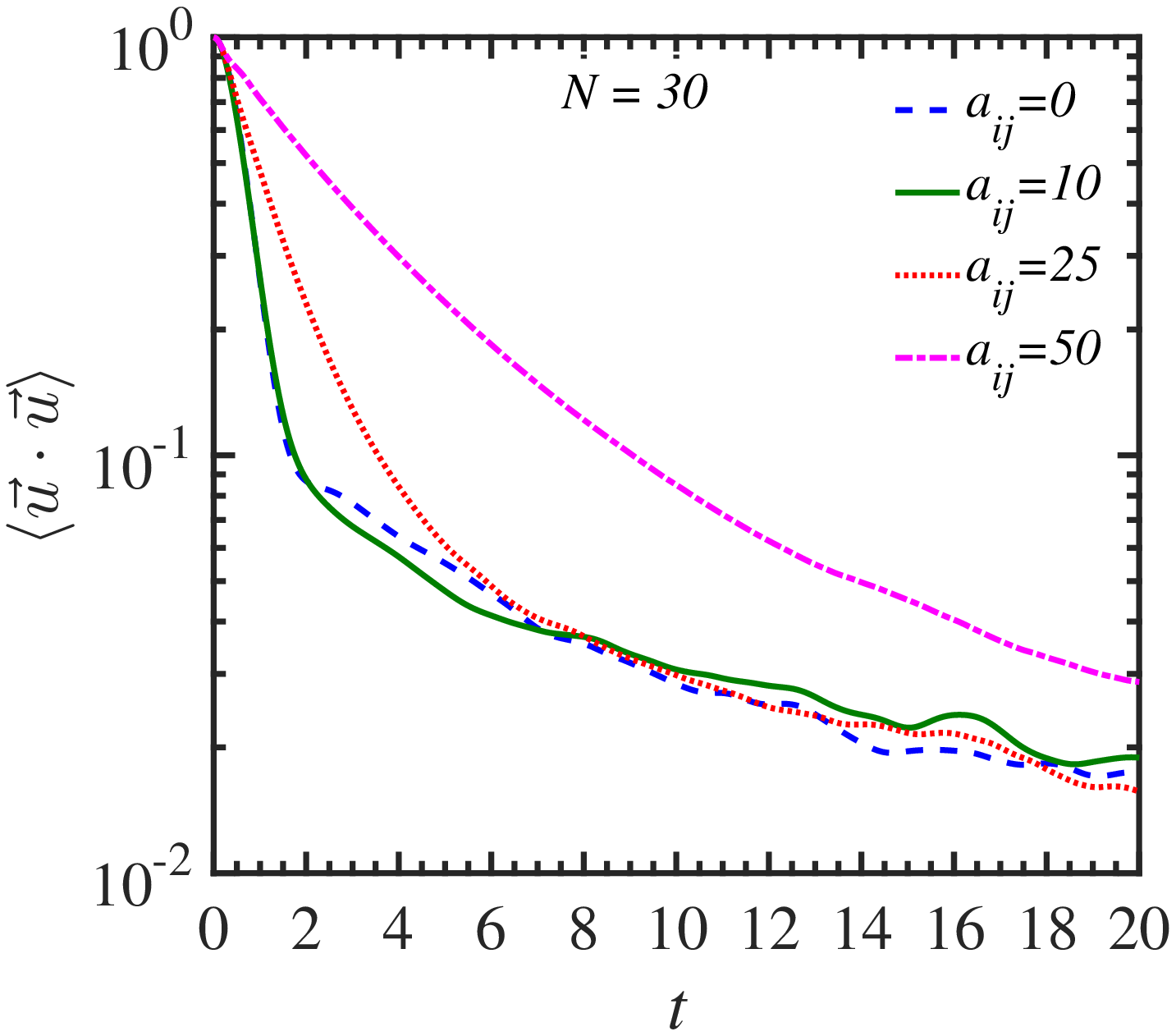}
  	\put (-160,160){($c$)}
  	\includegraphics[width=3in,height=2.7in]{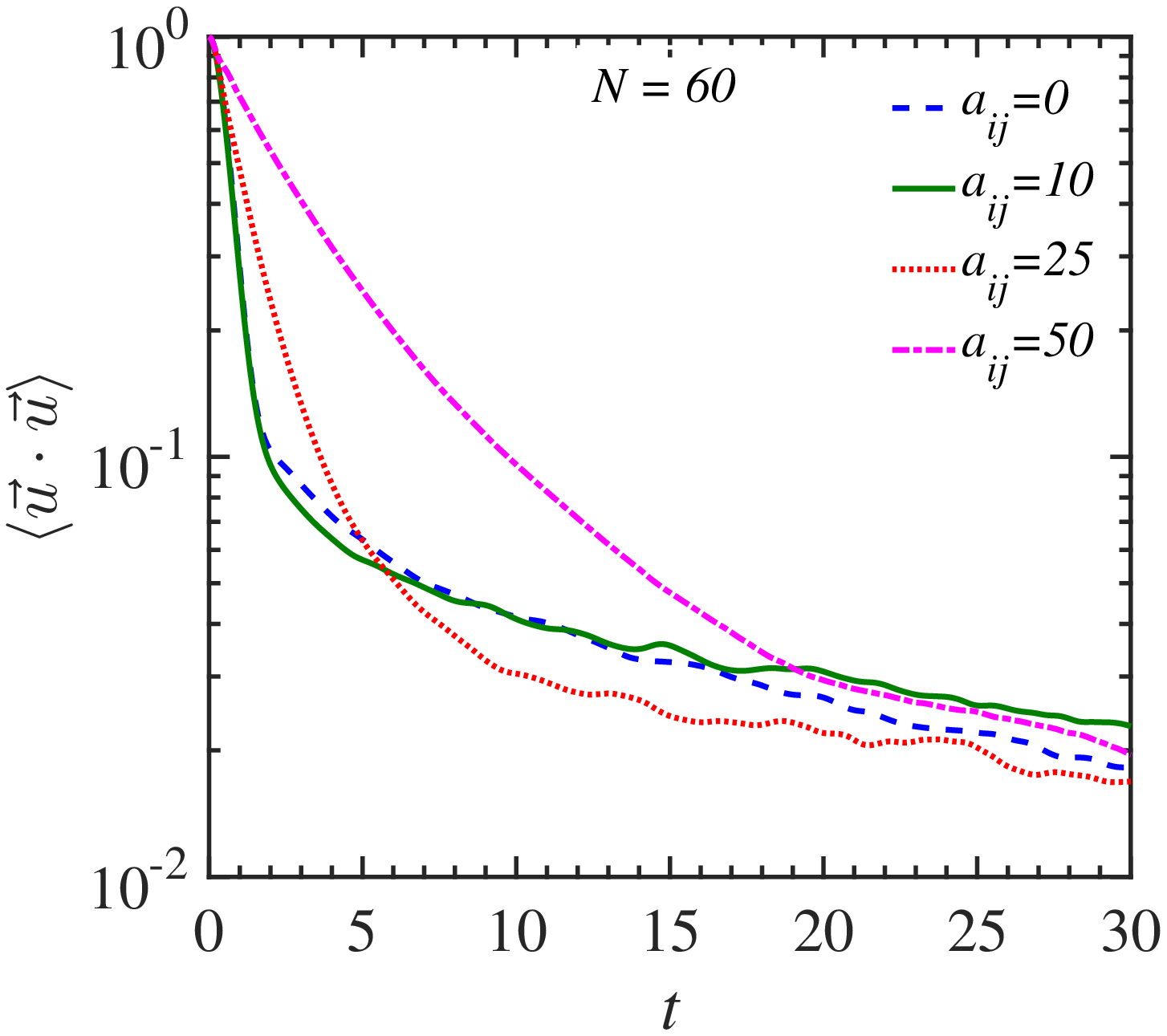}
  	\put (-160,160){($d$)}

  	  \caption {Relaxation dynamics of the backbone bonds, for chain lengths (\textit{a}) $N=10$, (\textit{b}) $N=20$, (\textit{c}) $N=30$ and (\textit{d}) $N=60$ for different values of repulsive parameter $a_{ij}=0$, $a_{ij}=10$, $a_{ij}=25$, and $a_{ij}=50$.  Note that, for $N=10$, the bond vector ACF is described well by single relaxation mode (appears as a straight line in semi-log plot)}
  \end{figure*}	

 \begin{figure*}
 	\includegraphics[width=2.6in,height=2.4in]{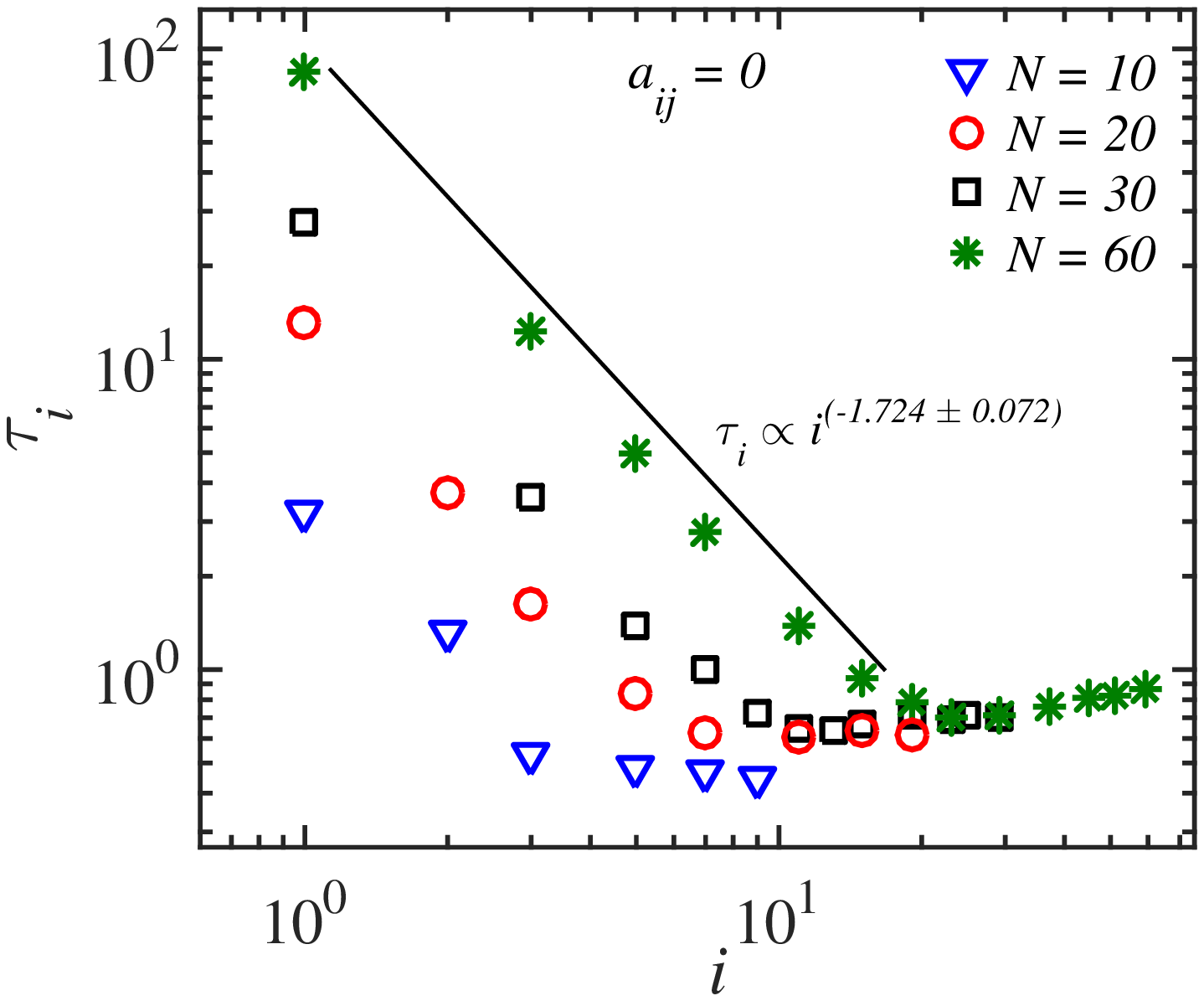}
 	\put (-150,35){($a$)}
 	\includegraphics[width=2.6in,height=2.4in]{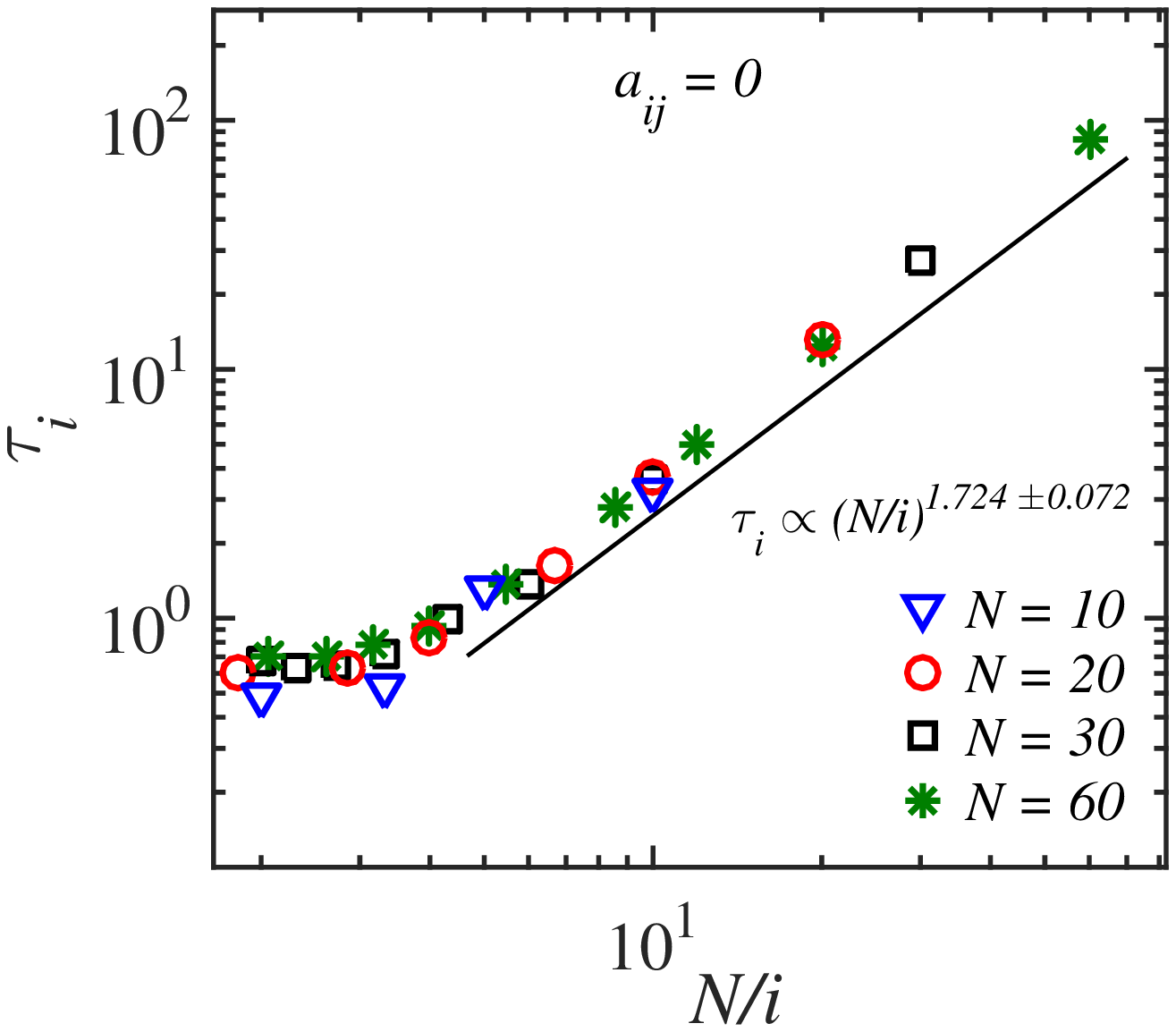}
 	\put (-150,145){($b$)}
 	
 	\includegraphics[width=2.6in,height=2.4in]{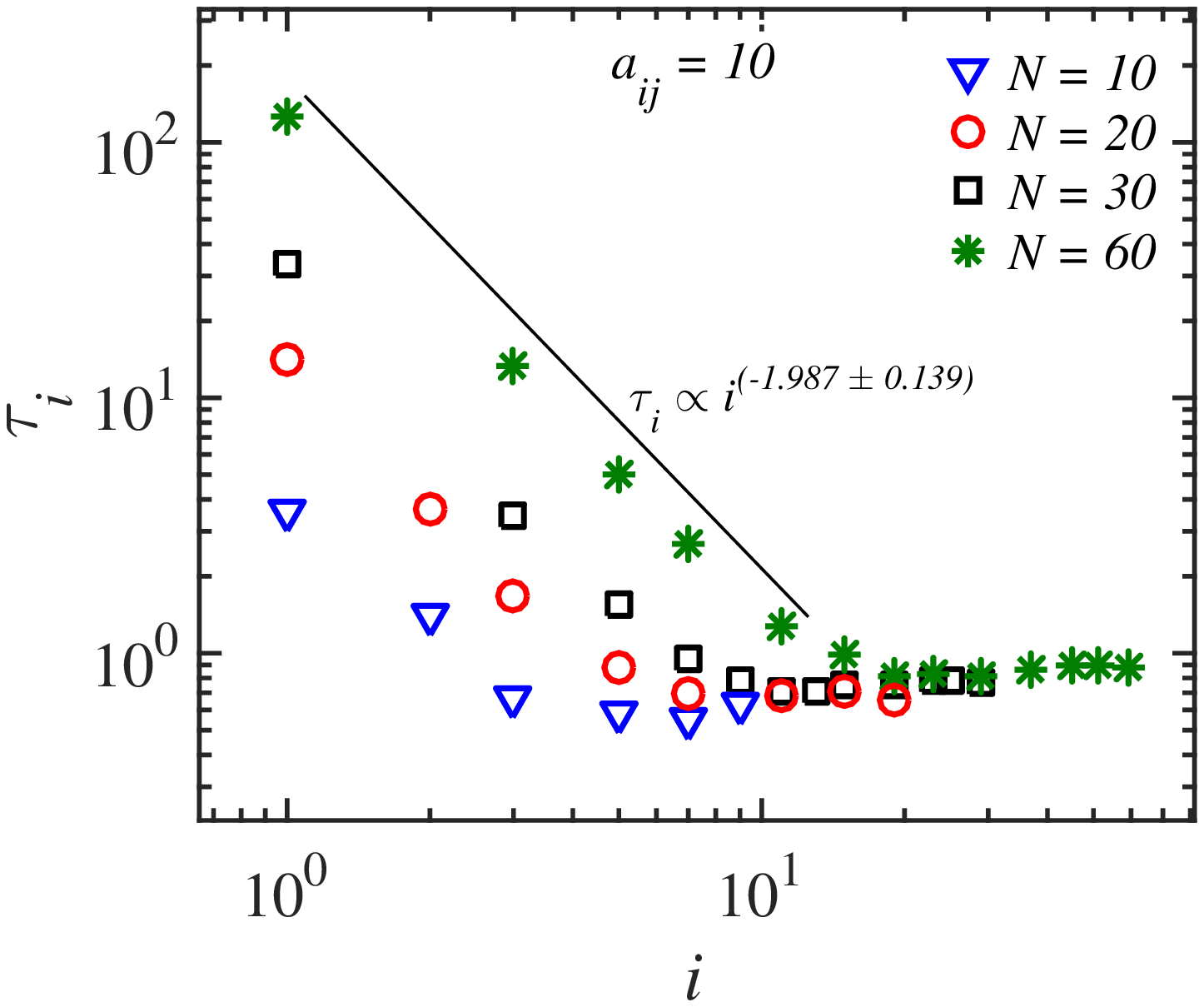}
 	\put (-150,40){($c$)}
 	\includegraphics[width=2.6in,height=2.4in]{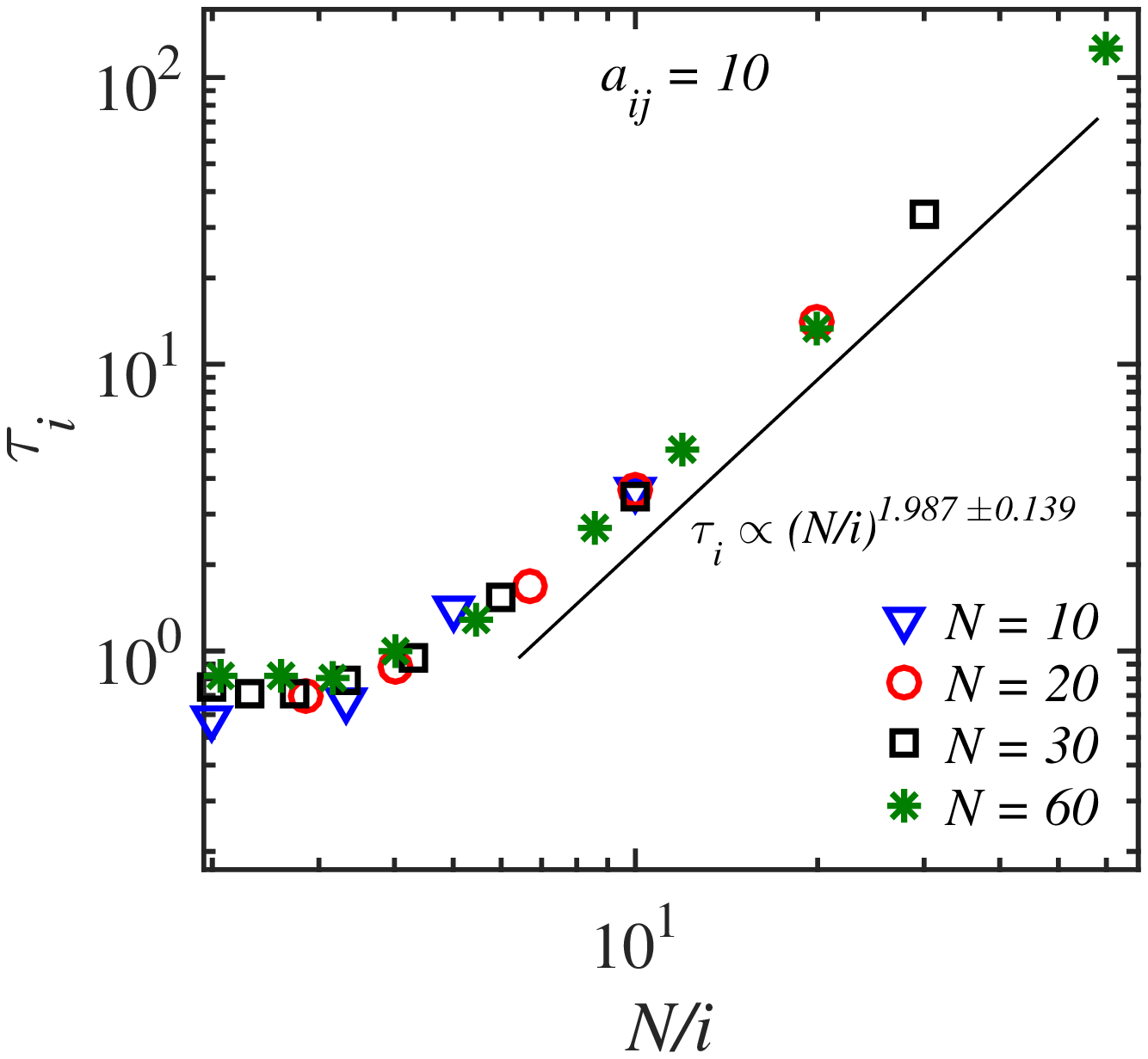}
 	\put (-150,145){($d$)}
 	
 	\includegraphics[width=2.6in,height=2.4in]{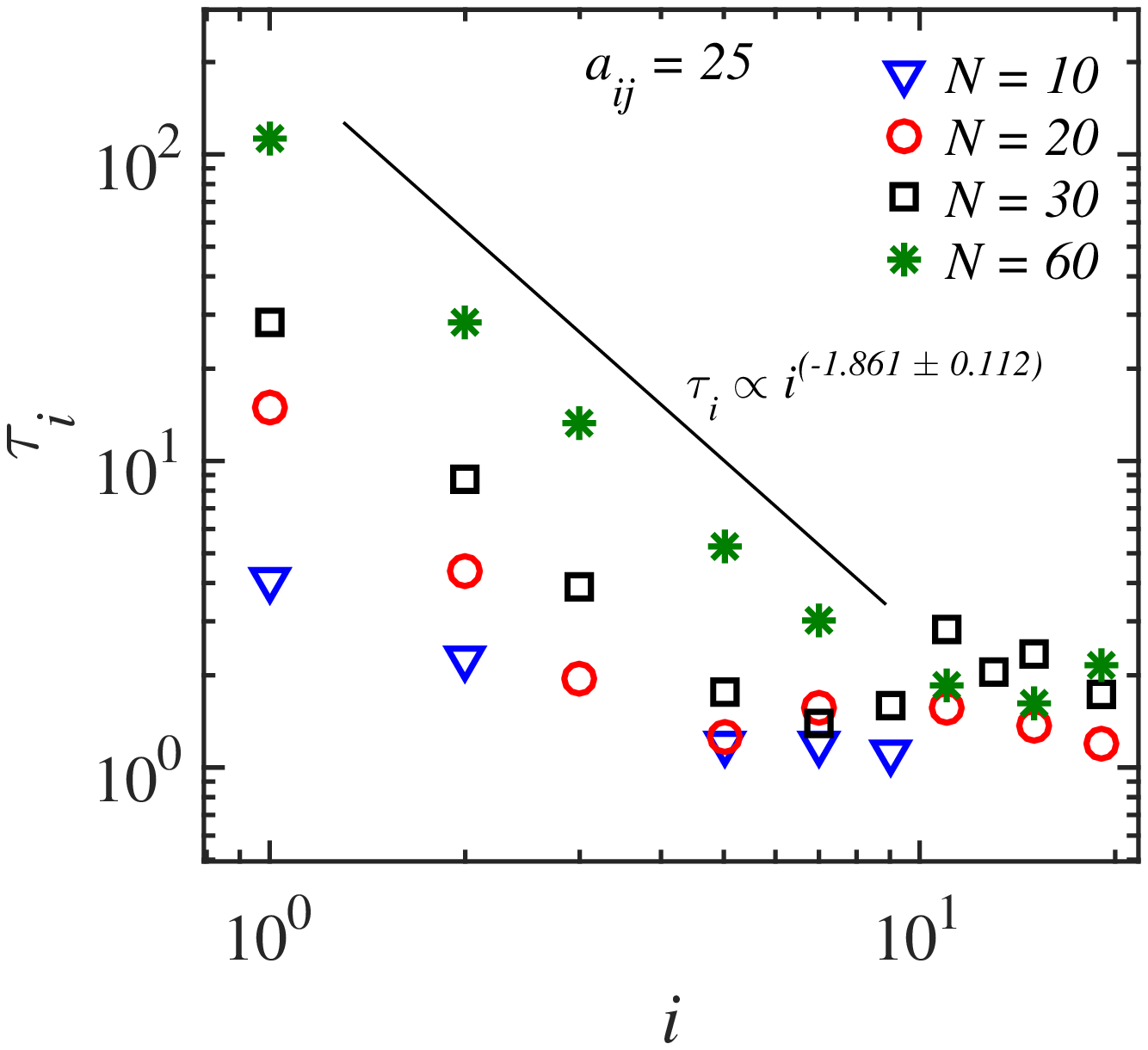}
 	\put (-145,40){($e$)}
 	\includegraphics[width=2.6in,height=2.4in]{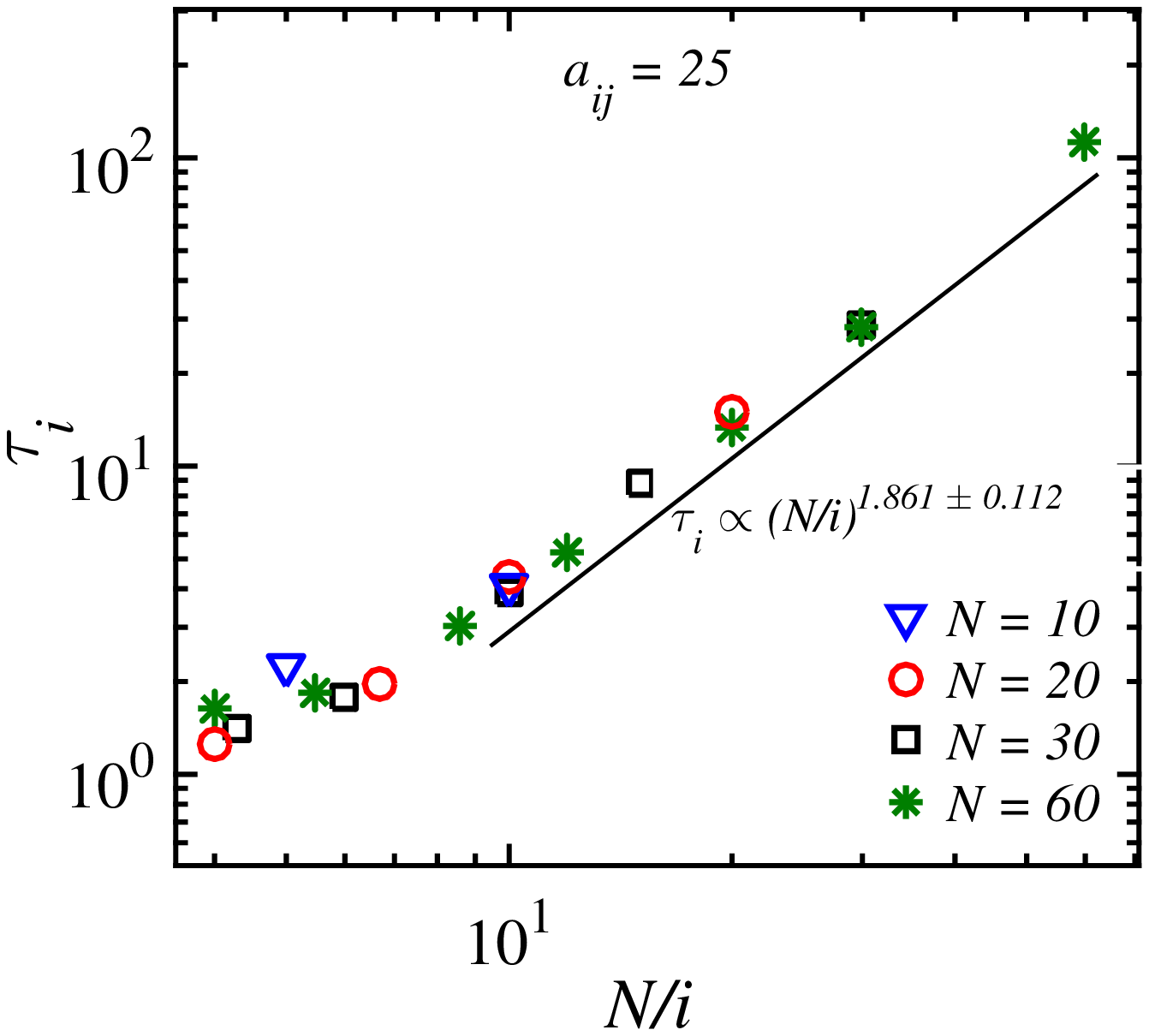}
 	\put (-150,145){($f$)}
 	
 	\caption {Relaxation times ($\tau_{i}$) of the $i^{th}$ mode versus $i$ (\textit{a, c, d}) and $N/i$ (\textit{b, d, f}), for $a_{ij}=0,10,$ and $25$ for different chain lengths. The solid line shows the power-law scaling of $\tau _i$ with $i$  and $N/i$.}
 \end{figure*}	
 
  \begin{figure*}
  	\includegraphics[width=3in,height=2.8in]{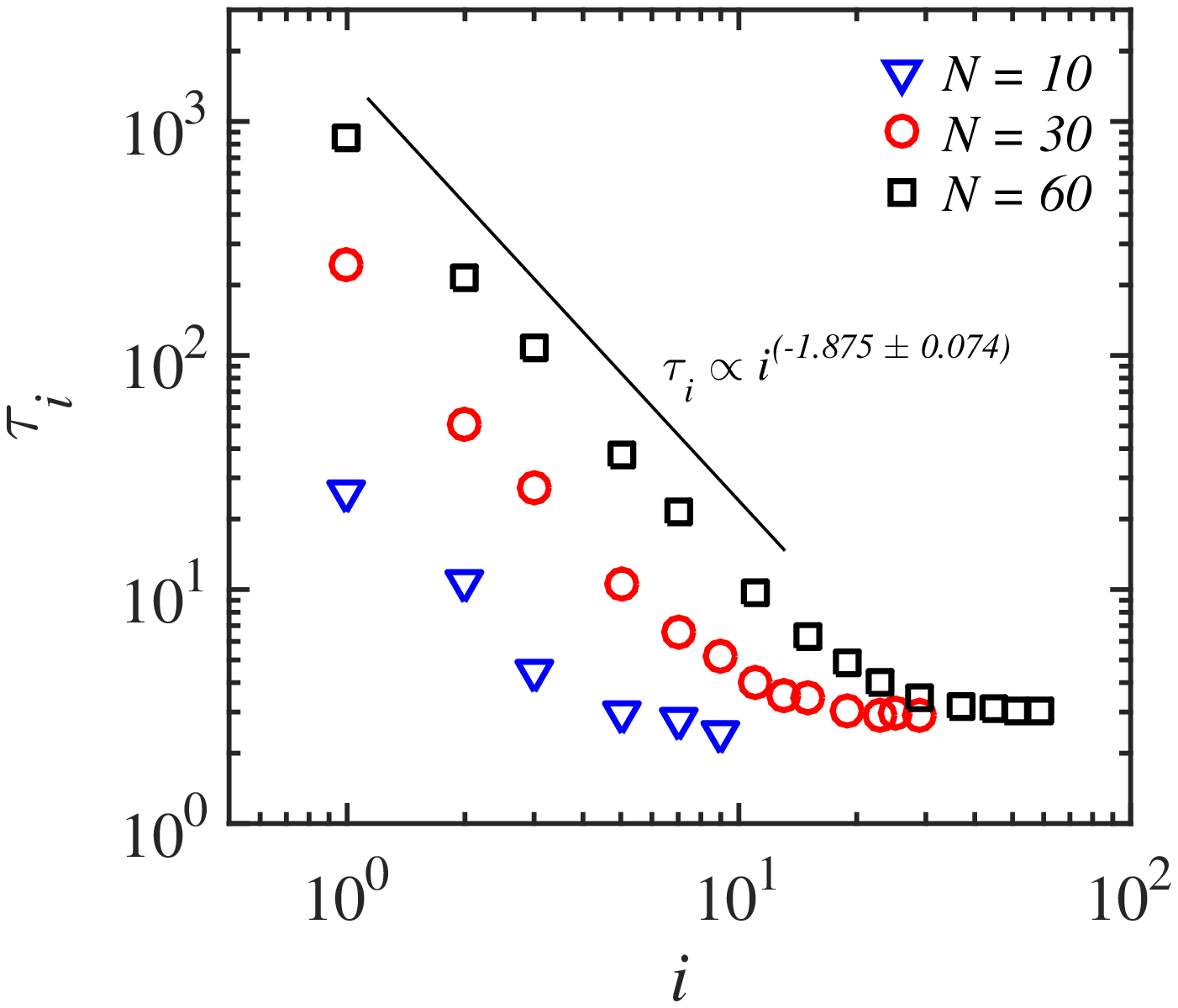}
  	\put (-110,170){($a$)}
  	\includegraphics[width=3in,height=2.8in]{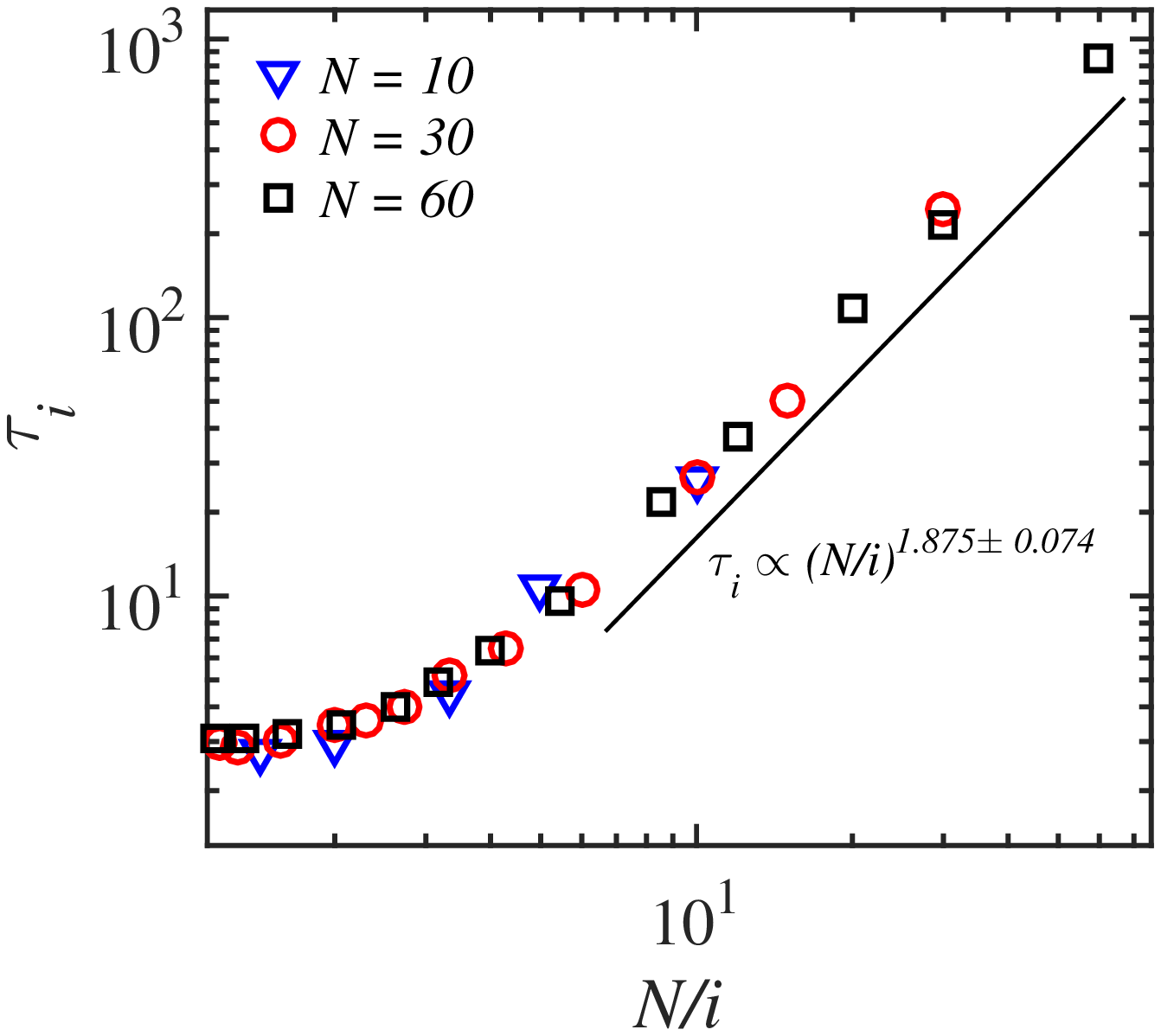}
  	\put (-110,170){($b$)}
  	
  	\includegraphics[width=3in,height=2.8in]{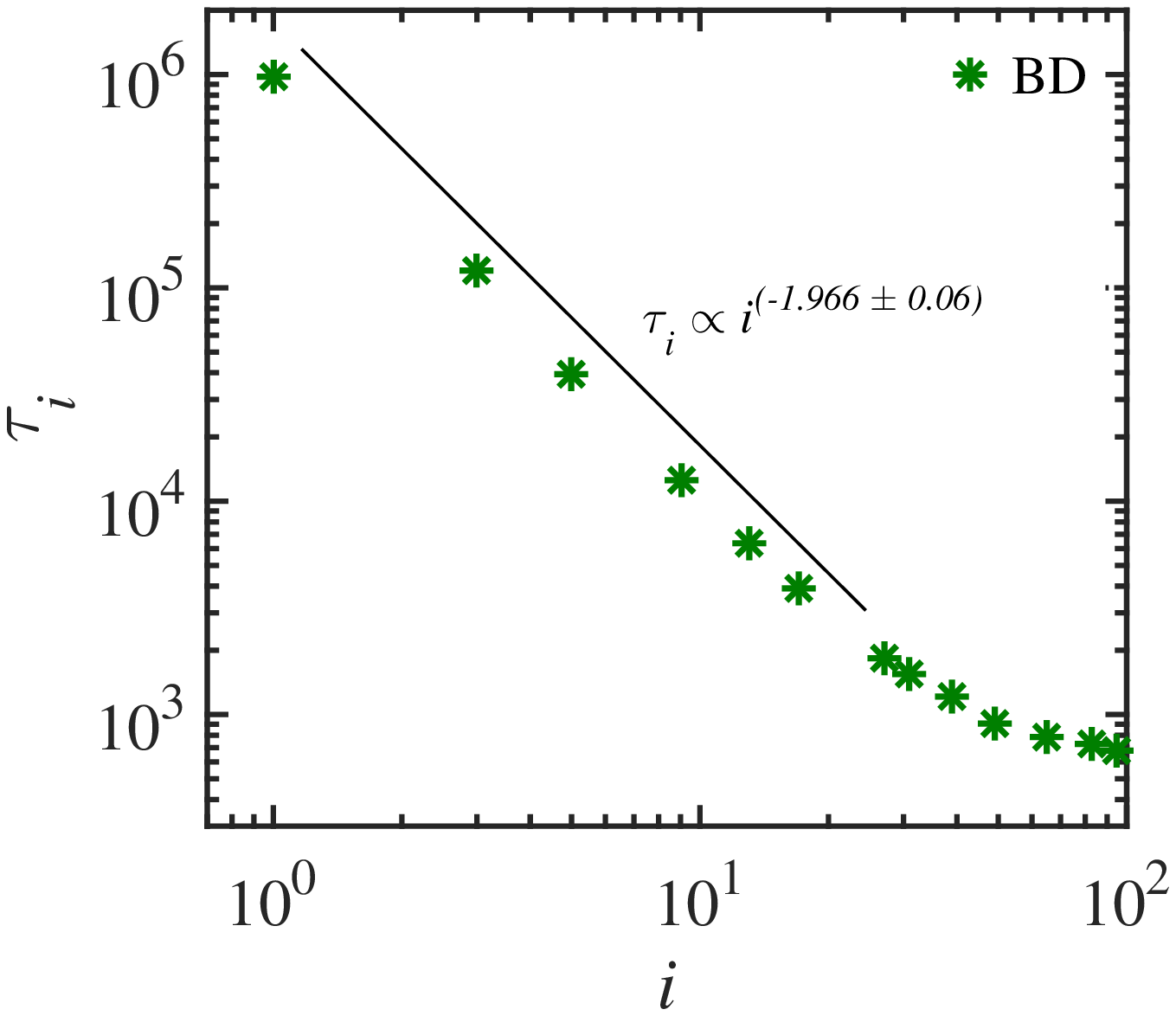}
  	\put (-110,170){($c$)}
  	\includegraphics[width=3in,height=2.8in]{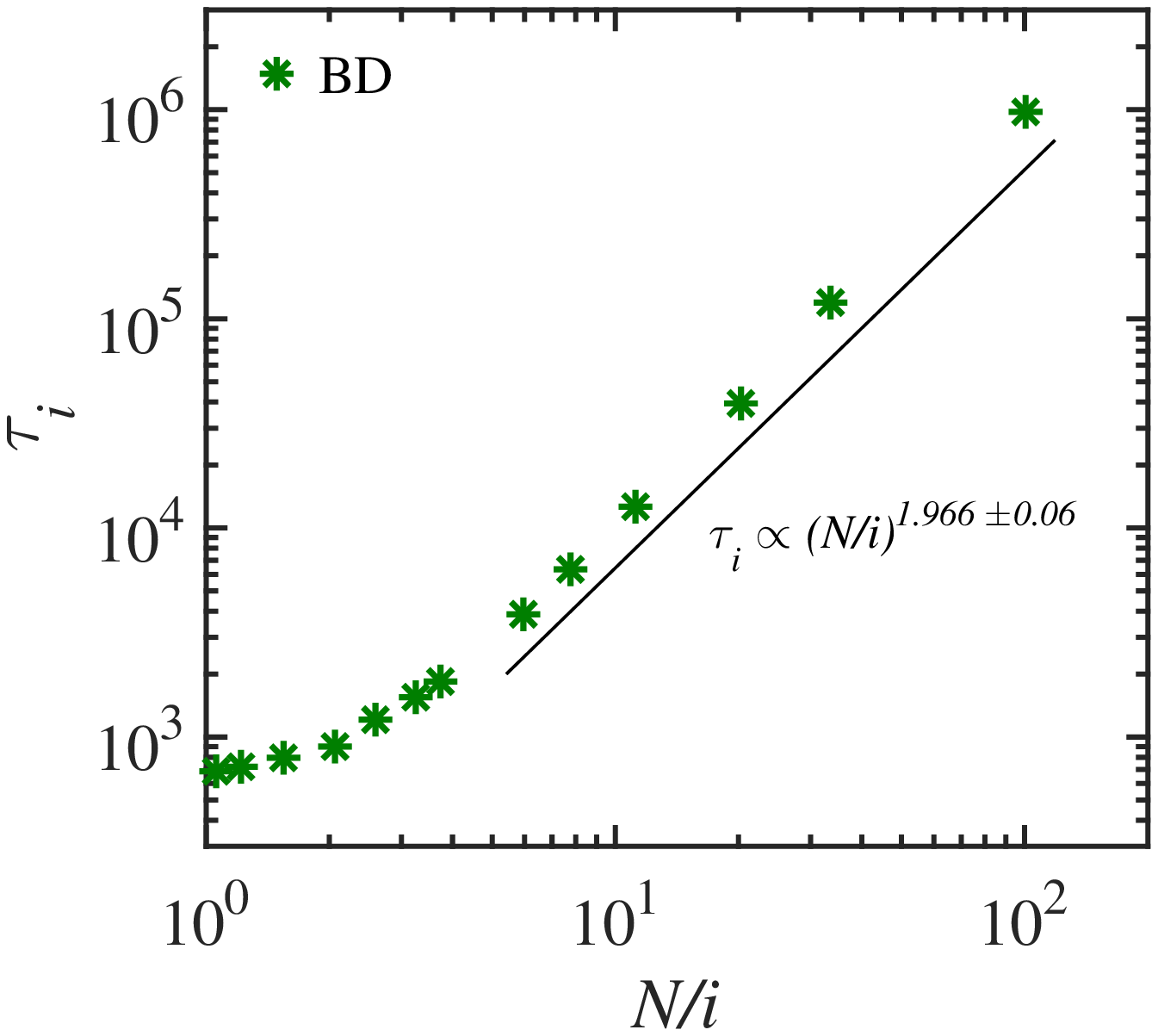}
  	\put (-110,170){($d$)}
  	
  	 	\caption {Variation of $\tau_i$ with $i$ (\textit{a, c}), and $N/i$ (\textit{b, d}) obtained from 
  	 		the Brownian dynamics (BD) simulations.}
  \end{figure*}

\begin{figure}
	\includegraphics[width=2.7in,height=2.5in]{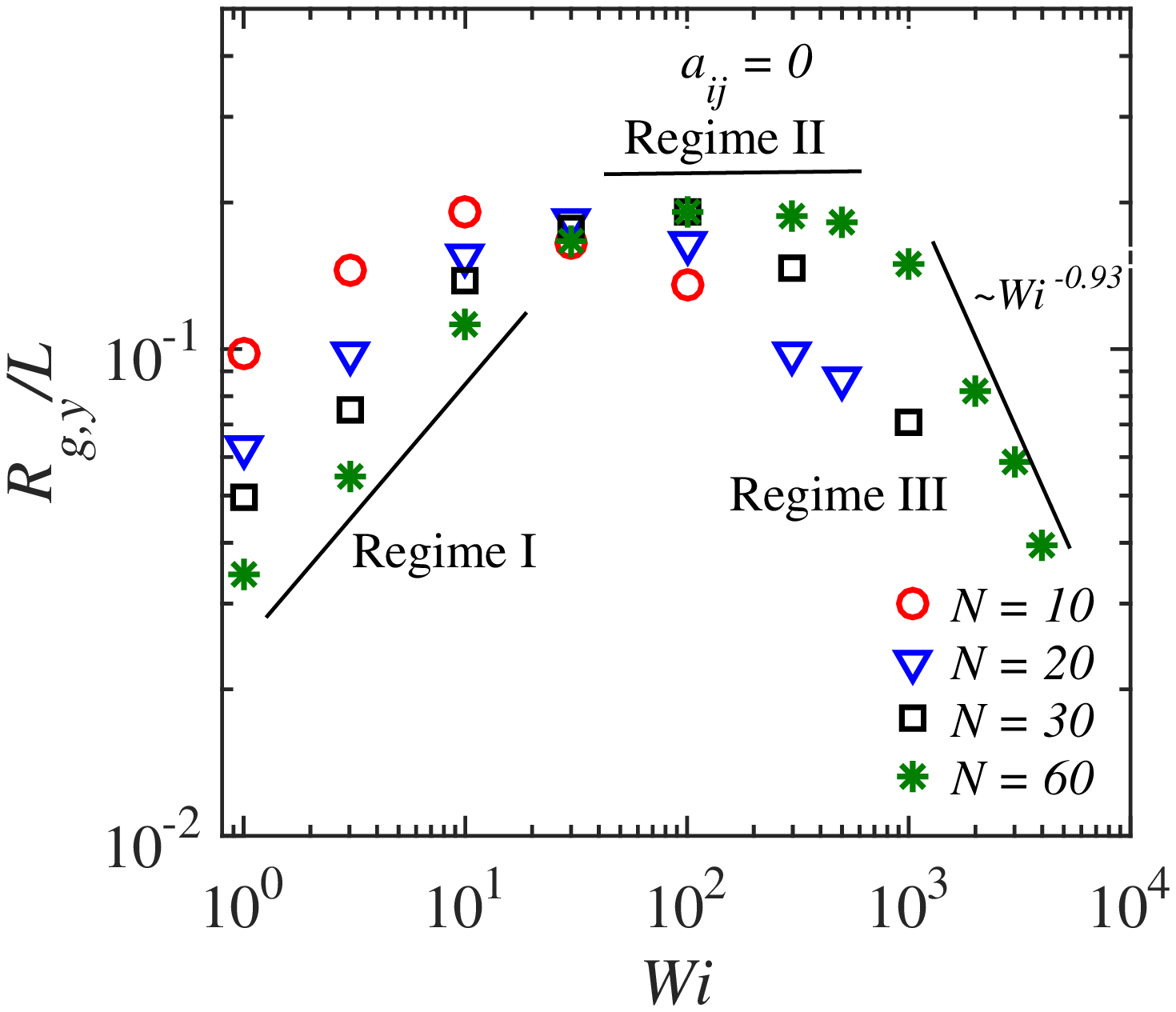}
	\put (-155,150){($a$)}
	\includegraphics[width=2.7in,height=2.5in]{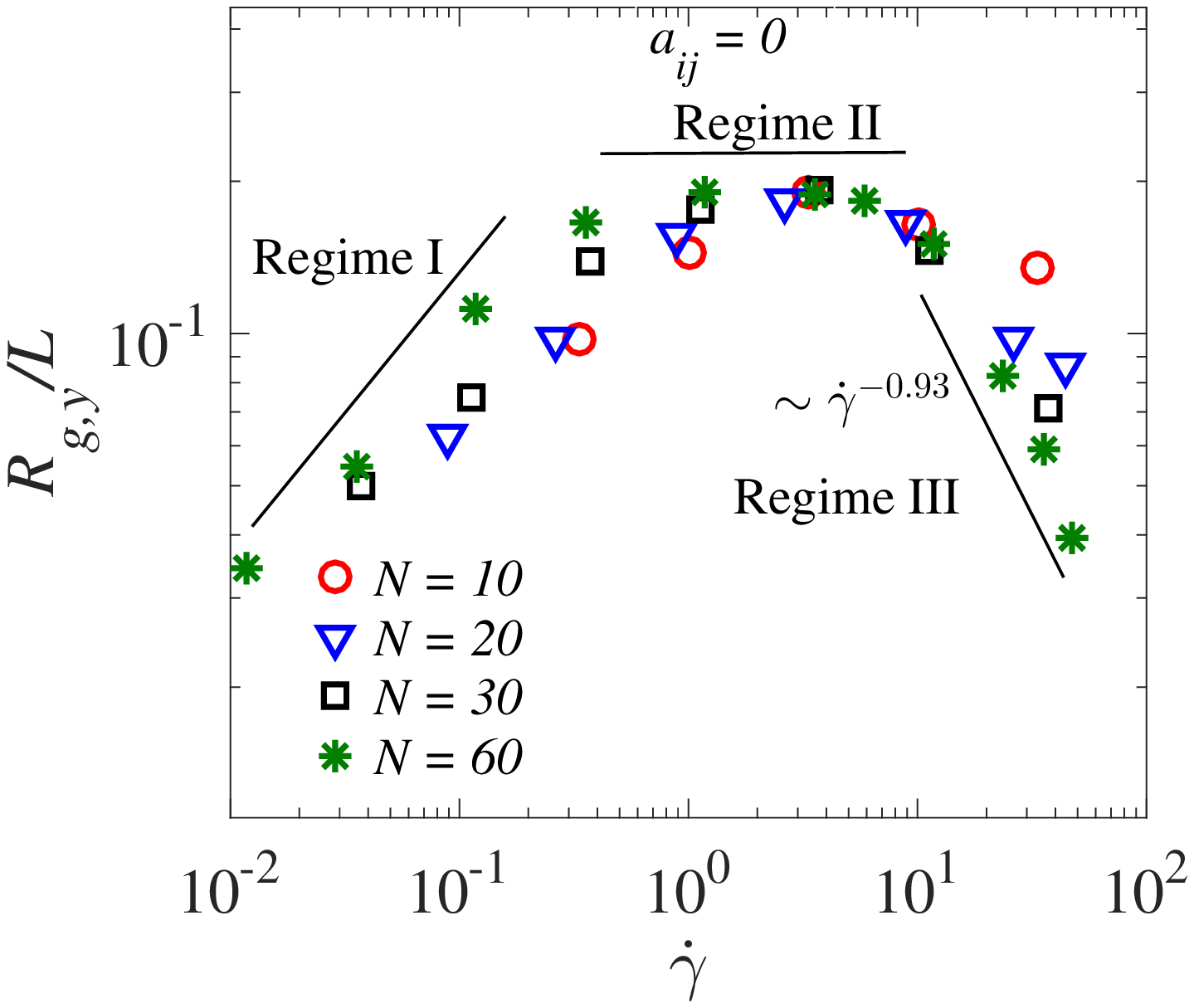}
	\put (-155,150){($d$)}

	\includegraphics[width=2.7in,height=2.5in]{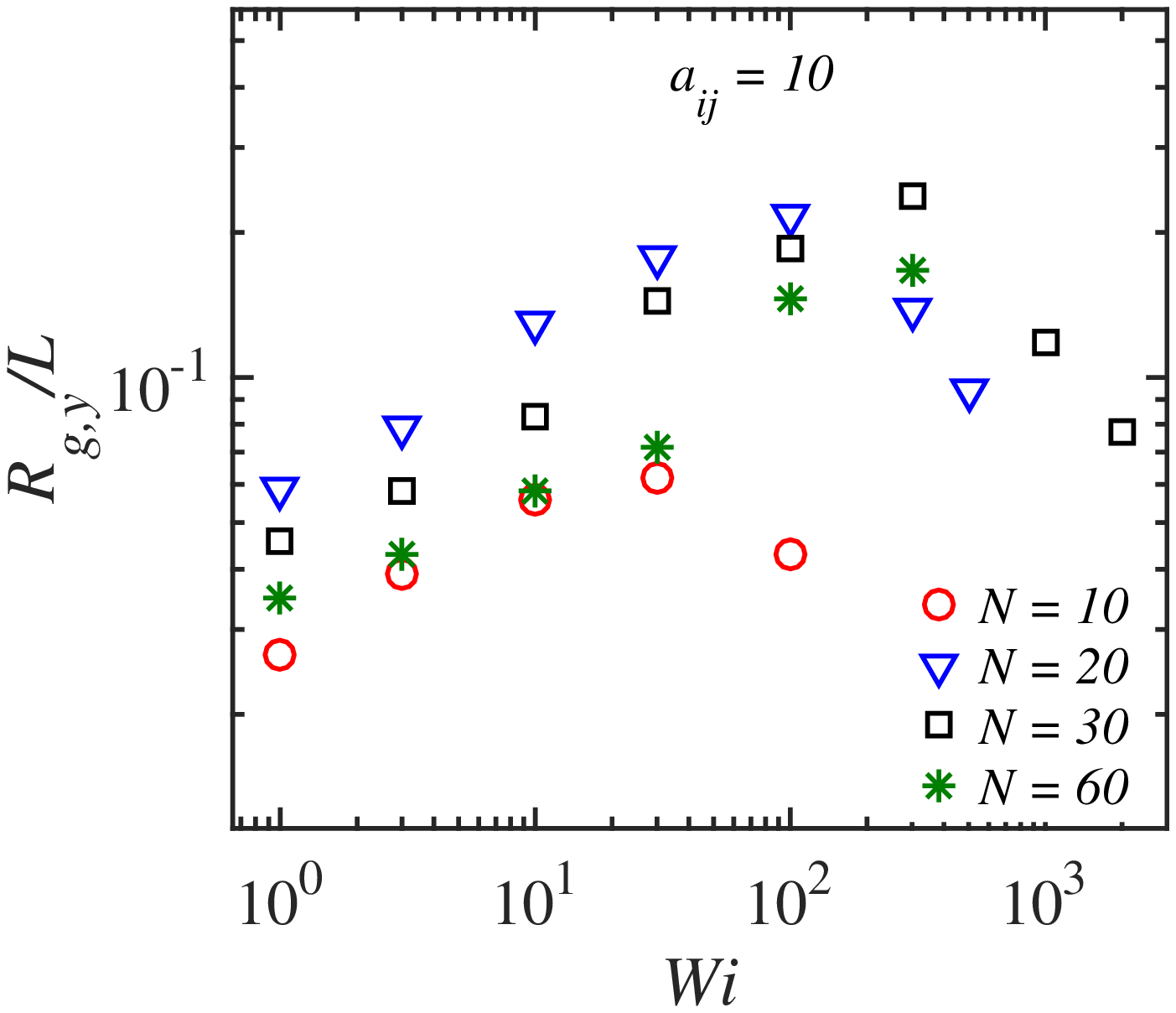}
	\put (-155,150){($b$)}
	\includegraphics[width=2.7in,height=2.5in]{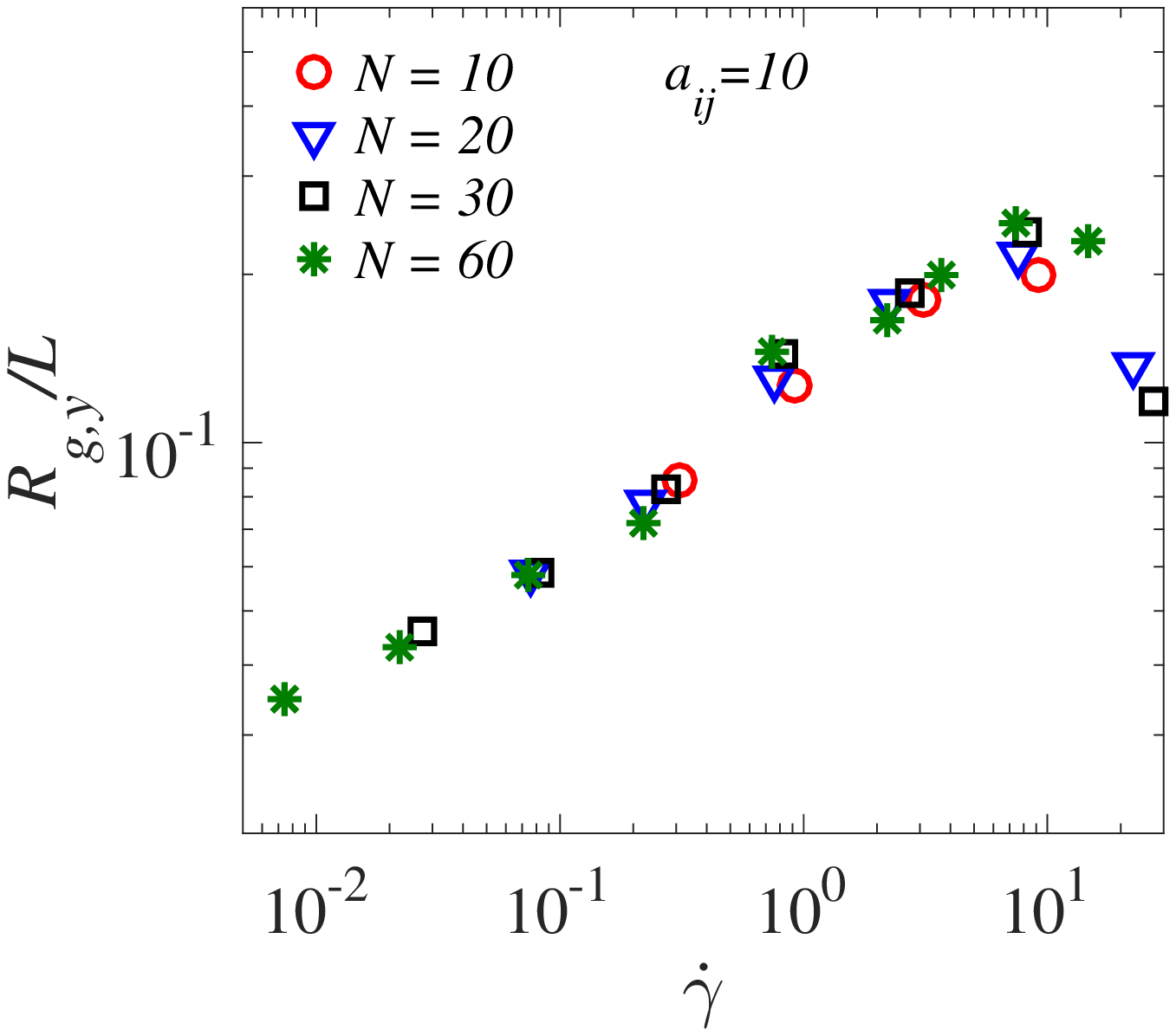}
	\put (-40,150){($e$)}	
	
	\includegraphics[width=2.7in,height=2.5in]{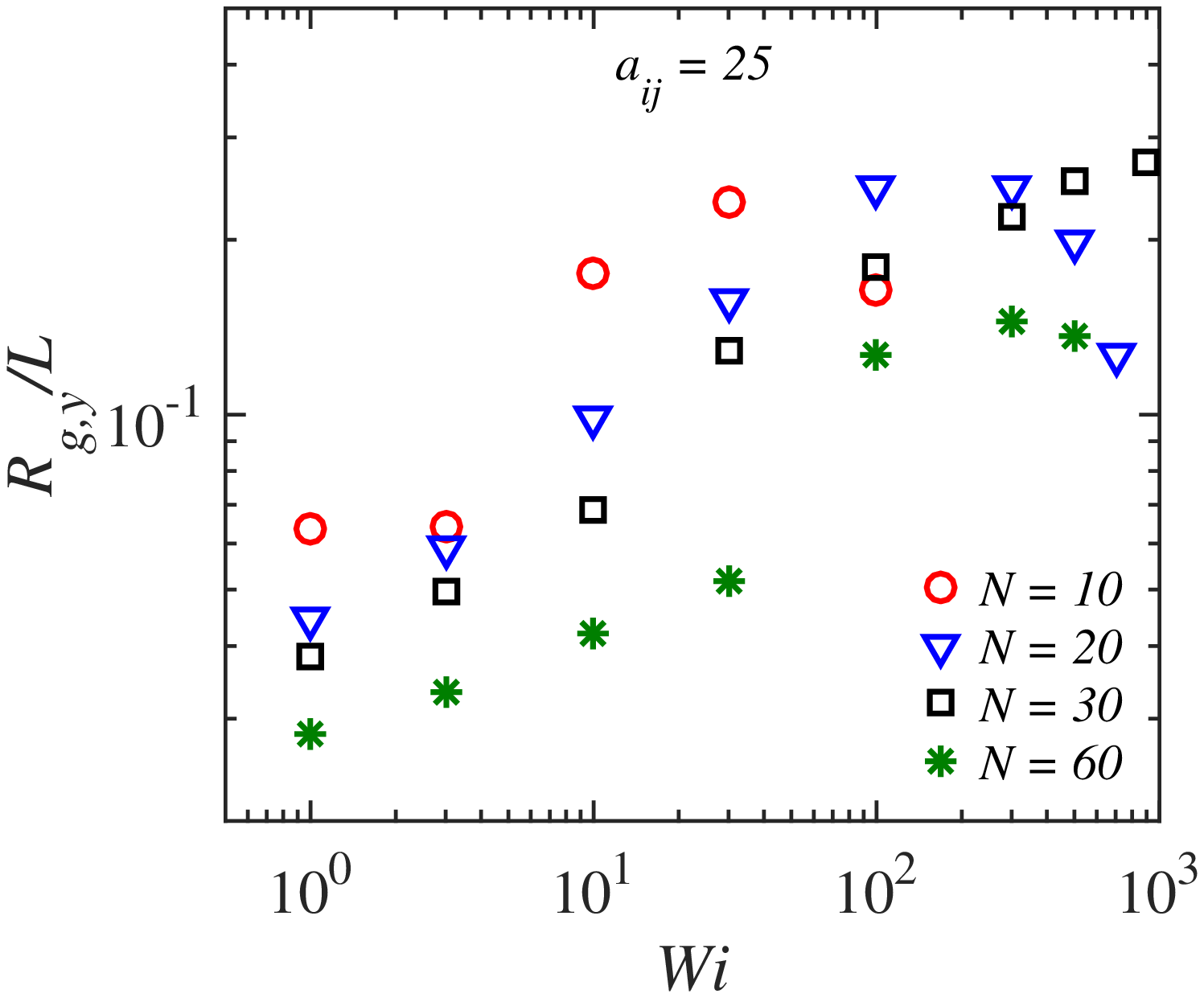}
	\put (-155,150){($c$)}
	\includegraphics[width=2.7in,height=2.5in]{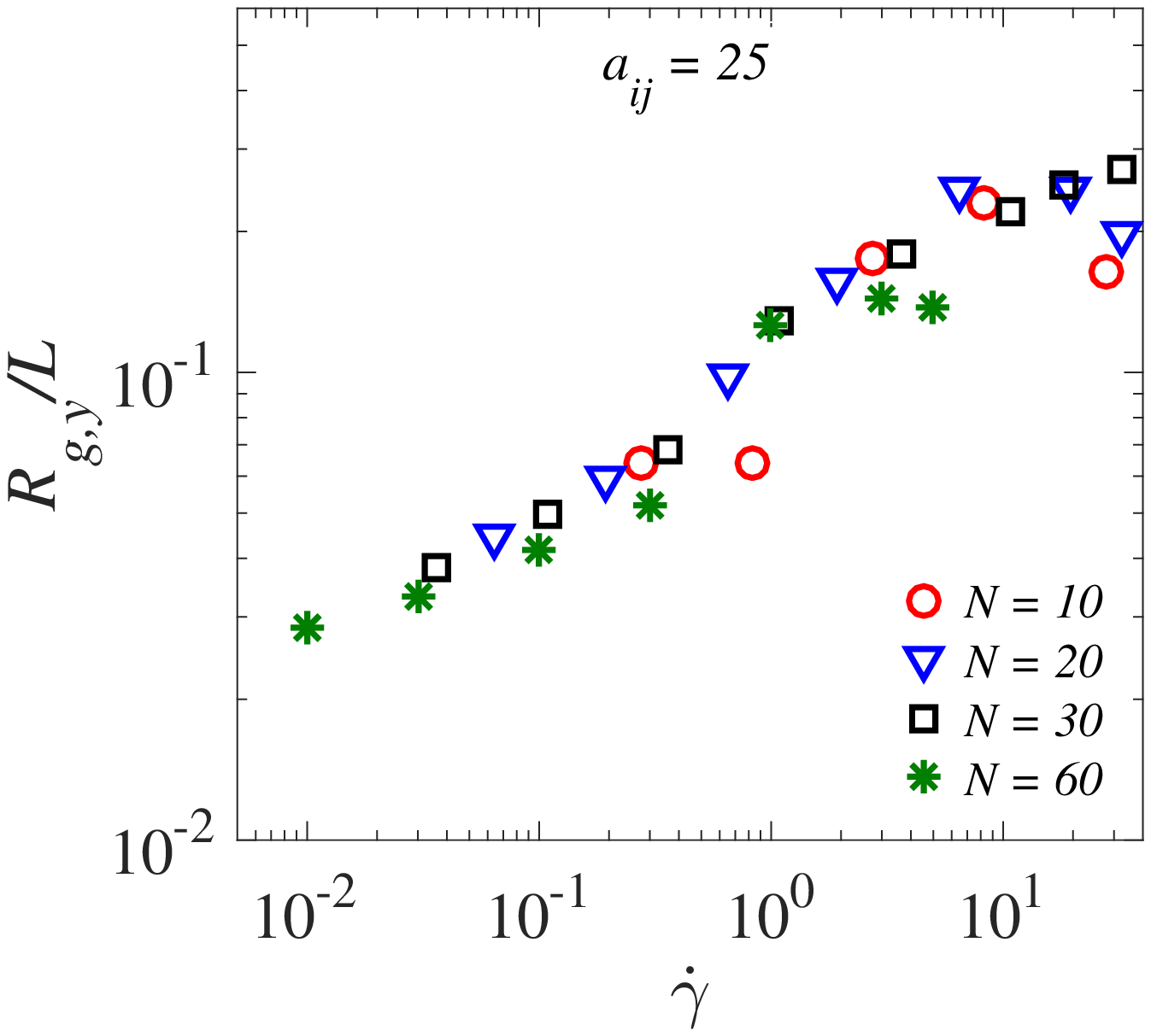}
	\put (-155,150){($f$)}	
	\caption { Variation of $R_{g,y}$ with shear rate: Figures (\textit{a-c}) show the variation with $Wi$ whereas (\textit{d-f}) show the same with shear rate.}
\end{figure}

 \begin{figure}
 	\includegraphics[width=3.2in,height=2.8in]{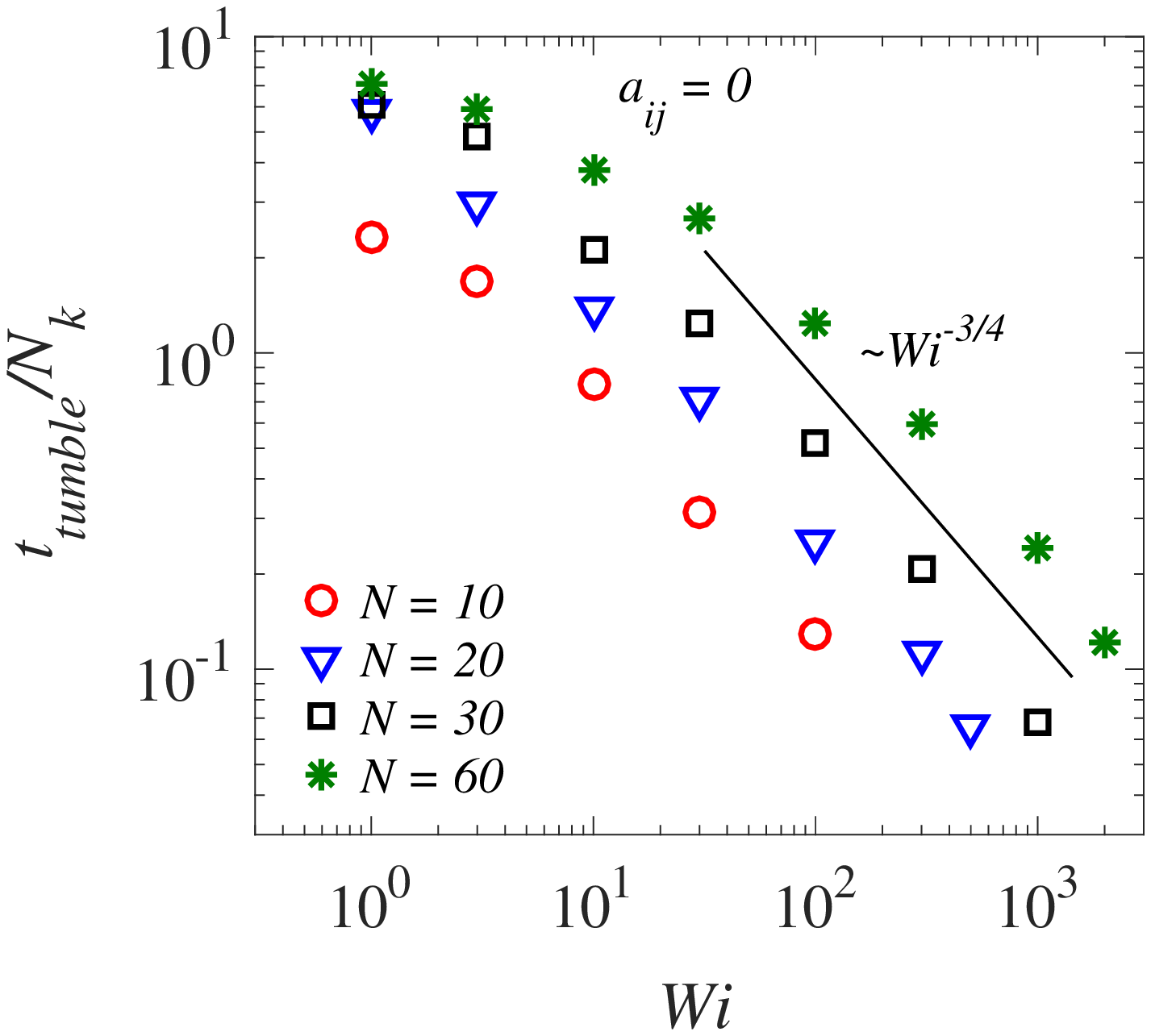}
 	\put (-45,170){($a$)}
 	
 	\includegraphics[width=3.2in,height=2.8in]{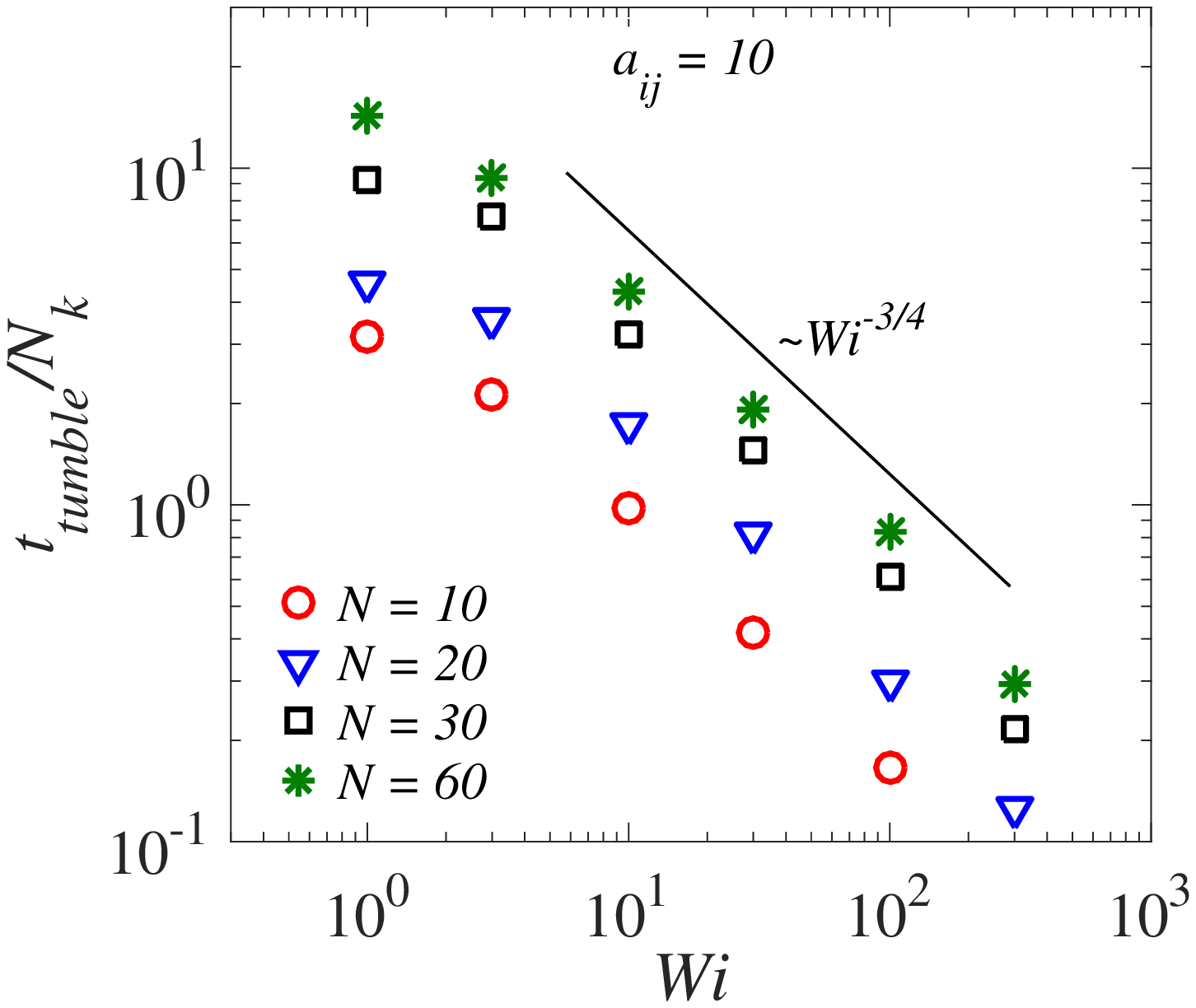}
 	\put (-45,170){($b$)}
 	
 	\includegraphics[width=3.2in,height=2.8in]{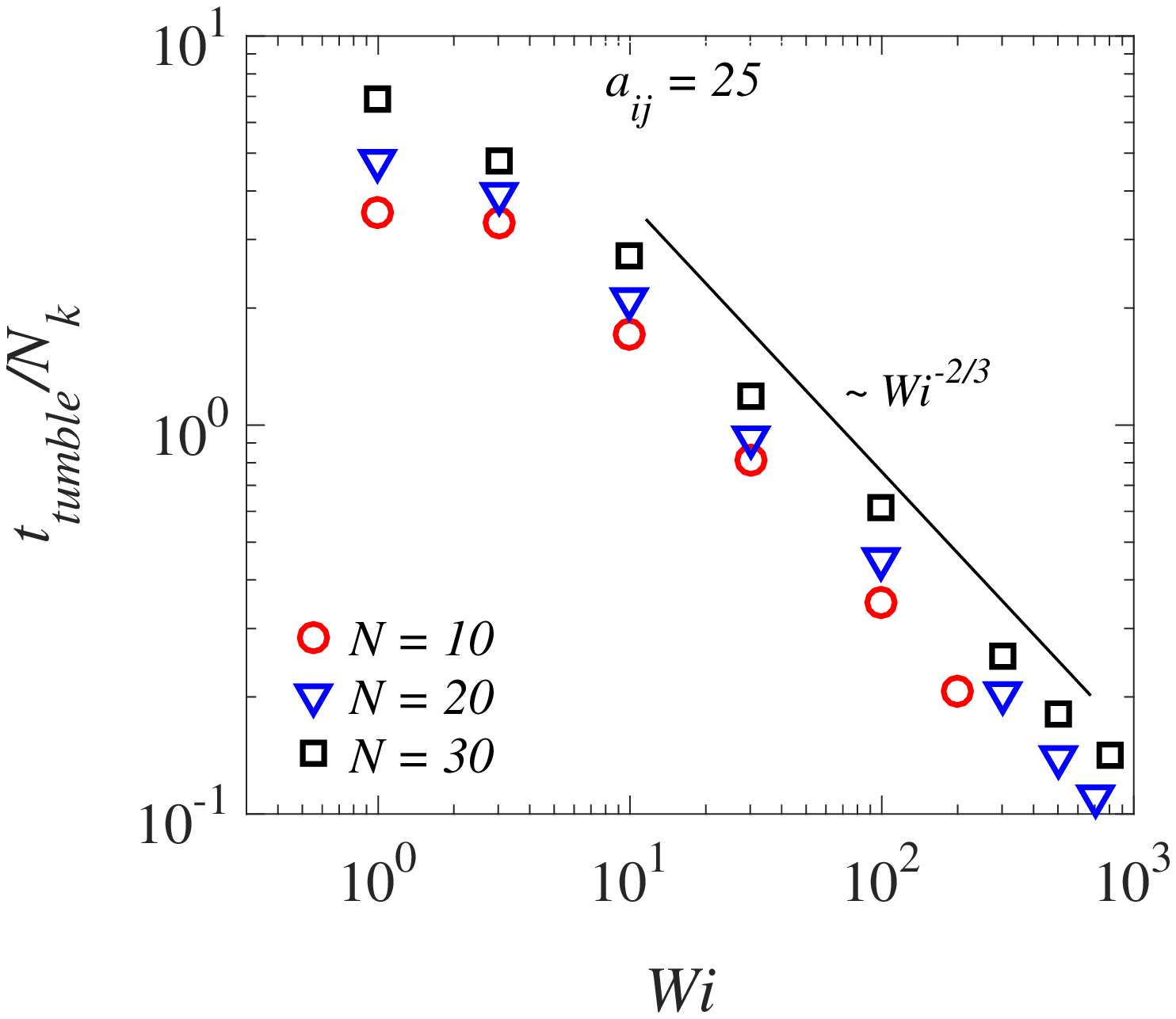}
 	\put (-45,170){($c$)}
	 \caption { Variation of the tumbling times (normalized by $N_k$) with $Wi$ for different values of the repulsive parameter $a_{ij}$. The solid line indicates the scaling law.}
 	
 \end{figure}

\end{document}